\documentclass[aps,prl,twocolumn,groupedaddress,nofootinbib,longbibliography]{revtex4-1}
\usepackage{amsmath}
\usepackage{amssymb}
\usepackage{graphicx,subfigure}
\usepackage{mathrsfs}
\usepackage[mathscr]{eucal}
\usepackage{ntheorem}
\usepackage[T1]{fontenc}
\usepackage{bm}
\usepackage{subfigure}
\usepackage[bookmarks=false]{hyperref}

\newcommand{\ket}[1]{|#1\rangle}
\newcommand{\bra}[1]{\langle #1|}

\newcommand{\expt}[1]{\langle #1 \rangle}

\newcommand{\tr}{\mathrm{tr}}

\newcommand{\abs}[1]{\lvert #1\rvert}
\newcommand{\norm}[1]{\lVert #1\rVert}

\def\CC{{\rm\kern.24em \vrule width.04em height1.46ex depth-.07ex \kern-.30em C}}
\def\RR{{\rm\kern.24em \vrule width.04em height1.46ex depth-.07ex
\kern-.30em R}}
\def\P{{\rm I\kern-.25em P}}

\DeclareMathAlphabet{\mathpzc}{OT1}{pzc}{m}{it}

\begin{document}

\title{Dissipative Quantum Metrology}
\author{Da-Jian Zhang}
\affiliation{Department of Physics, National University of Singapore, Singapore 117542}
\author{Jiangbin Gong}
\email{phygj@nus.edu.sg}
\affiliation{Department of Physics, National University of Singapore, Singapore 117542}

\date{\today}

\begin{abstract}
Conventional strategies of quantum metrology are built upon POVMs, thereby possessing several general features, including the demolition of the state to be measured, the need of performing a number of measurements, and the degradation of performance under decoherence and dissipation. Here, we propose an innovative measurement scheme, called dissipative adiabatic measurements (DAMs), based on which, we further develop an approach to estimation of parameters characterizing dissipative processes. Unlike a POVM, whose outcome is one of the eigenvalues of an observable, a DAM yields the expectation value of the observable as its outcome, without collapsing the state to be measured. By virtue of the very nature of DAMs, our approach is capable of solving the estimation problem in a state-protective fashion  with only $M$ measurements, where $M$ is the number of parameters to be estimated. More importantly, contrary to the common wisdom, it embraces decoherence and dissipation as beneficial effects and offers a Heisenberg-like scaling of precision, thus outperforming conventional strategies. Our DAM-based approach is direct, efficient, and expected to be immensely useful in the context of dissipative quantum information processing.
\end{abstract}

\maketitle

Estimation of unknown parameters characterizing dynamical processes is a pivotal task throughout quantum science and technologies \cite{2009Paris125}.
Conventional strategies of quantum metrology \cite{2006Giovannetti10401} are to let many probes evolve under the process in question and perform positive operator-valued measures (POVMs) on the evolved states. As the outcome of a POVM is nothing but an individual eigenvalue of an observable, which is not directly related to the parameters of interest, the values of these parameters have to be inferred indirectly from a classical post-processing of the measurement outcomes. Naturally, the goal is to estimate the parameters as precisely as possible with given resources. Yet, due to the indirectness of conventional strategies, there is an inherent statistical error in any such estimation, which cannot be avoided even in optimal conditions. Indeed, in the case of the probes being uncorrelated, the quantum Cram\'{e}r-Rao inequality \cite{1994Braunstein3439} imposes an ultimate bound on precision, which is known as the standard quantum limit, stating that the error on average at least scales as $1/\sqrt{N}$. Here, $N$ represents the amount of resources employed in the estimation procedure, which is often referred to as the number of POVMs performed but can be other kinds of quantities, e.g., the evolution time of the probe undergoing the process \cite{2006Giovannetti10401}. By exploiting quantum effects, such as entanglement and squeezing, the precision can be further improved to the Heisenberg scaling $1/N$ for unitary dynamics \cite{2004Giovannetti1330,2011Giovannetti222}. Such a quadratic improvement is, however, typically elusive in the presence of decoherence and dissipation \cite{1997Huelga3865,2007Monras160401,2009Dorner40403a,2011Escher406,
2012Demkowicz-Dobrzanski1063,2014Alipour120405}, due to the fragility of quantum effects under their influence.

In this Letter, targeting at parameter estimation in dissipative processes, we develop an innovative approach going beyond conventional  strategies.   This is particularly timely in view of the recent advent of dissipative quantum information processing protocols, such as quantum state preparation \cite{2008Diehl878,2011Kastoryano90502,2011Cho20504,2011Krauter80503,
2011Vollbrecht120502,2013Carr33607,2013Torre120402,2013Rao33606,
2014Bentley40501,2016Abdi233604,2016Kimchi-Schwartz240503,2016Reiter40501,
2016Znidaric30403}, quantum computation \cite{2009Verstraete633,2013Kastoryano110501}, and quantum
simulation \cite{2010Weimer382,2011Barreiro486}.
The working principle of these dissipation-based protocols is to finely tune controllable parameters of a dissipative system such that its irreversible dynamics drives the system into a desired steady state, regardless of its initial state. So, the estimation problem addressed in this Letter is precisely the converse, i.e., determining these parameters through measuring the steady state, which is extremely relevant to these protocols.
Unfortunately, it seems daunting to use conventional strategies to solve this problem, for the following reasons. First, for dissipative processes considered here, these strategies generally provide no substantial improvement over classical strategies on precision. Second, steady states in the above protocols are typically entangled states or some other desirable states. One is thus unwilling to demolish them, which is, however, unavoidable for conventional strategies, due to the use of POVMs. Third, performing a number of measurements is necessary in conventional strategies. Yet, this costs not only many copies of steady states but also a lot of experimental effort.

\begin{figure}[htbp]
\includegraphics[width=0.4\textwidth]{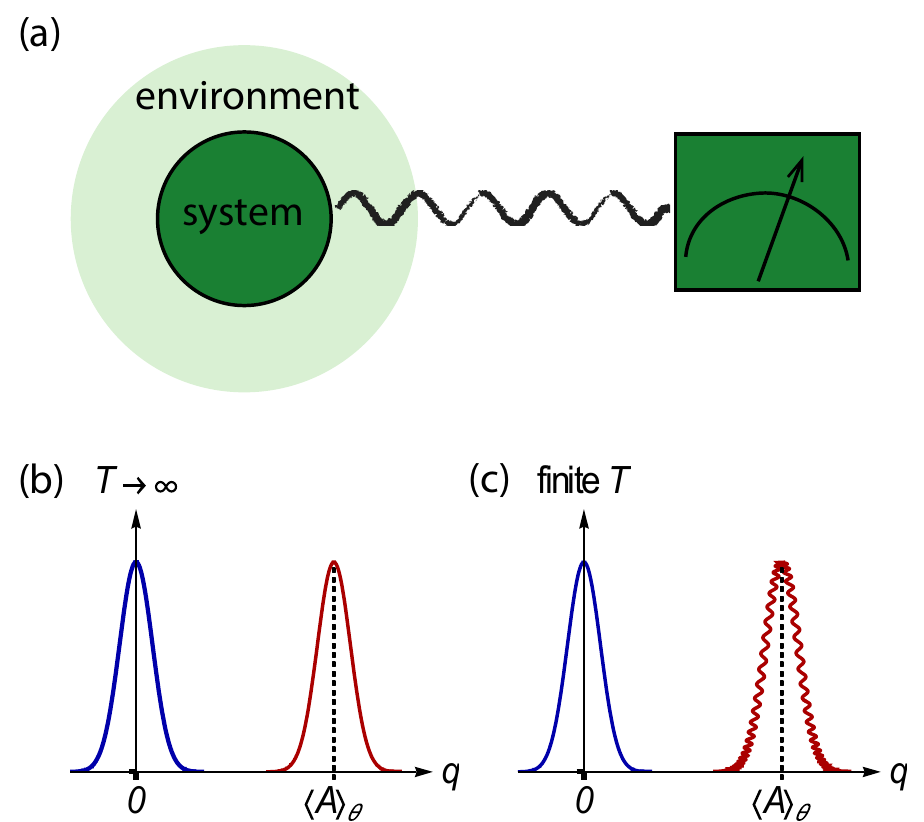}
\caption{Schematic representation of DAMs. (a) Setup: a dissipative system coupled to a measuring apparatus via an extremely weak but long-time interaction, with coupling strength $T^{-1}$ and coupling time $T$. (b) The apparatus is initially prepared in a Gaussian centered at $q=0$ (represented by the blue curve), which, in the limit of $T\rightarrow\infty$, evolves into the Gaussian centered at $q=\expt{A}_\theta$ (represented by the red curve). (c) In a finite-time evolution, non-adiabatic effects exist, causing slight deviations of the evolved state from the ideal Gaussian.}
\label{fig1}
\end{figure}

Here, inspired by Aharonov \textit{et al.}'s adiabatic measurements (AAMs) \cite{1993Aharonov38}, we propose a new scheme of measurements tailored for dissipative systems.  We call such measurements  dissipative adiabatic measurements (DAMs). The system to be measured in a DAM is a dissipative system, coupled to a measuring apparatus via an extremely weak but long-time interaction (see Fig.~\ref{fig1}a).
The dynamics of the measuring procedure is dominated by the dissipative process, which continuously projects the system into its steady state. Such a dissipation-induced ``quantum Zeno effect'' effectively decouples the system from the apparatus in the long-time limit, eliminating the so-called quantum back action of measurements \cite{2004Giovannetti1330}.
Unlike POVMs, a DAM therefore does not collapse the state to be measured.
Moreover, as shown below, its outcome is the expectation value of an observable, which is directly related to the parameters of interest.
By utilizing the very nature of DAMs, our approach is able to solve the estimation problem in a state-protective fashion with only $M$ measurements, where $M$ is the number of parameters. In particular, in the single-parameter case, only one measurement is needed. Interestingly, contrary to the common wisdom, decoherence and dissipation are no longer undesirable but play an integral part in our approach. More importantly, our approach offers a Heisenberg-like scaling of precision, thus outperforming conventional strategies.

Let us start with a simple example. Consider the generalized amplitude damping process \cite{2010Nielsen} described by the Lindblad equation, $\frac{d}{d t}\rho(t)=\mathcal{L}_\theta\rho(t)$, where
$\mathcal{L}_\theta\rho:=\theta\left(\sigma_{-}\rho\sigma_{+}-\frac{1}{2}\{\sigma_{+}
\sigma_{-},\rho\}\right)
+(1-\theta)\left(\sigma_{+}\rho\sigma_{-}-\frac{1}{2}\{\sigma_{-}
\sigma_{+},\rho\}\right)$ is a Liouvillian superoperator depending on the parameter $\theta\in(0,1)$, with $\sigma_-=\ket{0}\bra{1}$ and $\sigma_+=\ket{1}\bra{0}$.
For this dissipative process, there is a unique steady state  $\rho_\theta:=\textrm{diag}(\theta,1-\theta)$, which can be approached if the evolution time is much longer than the relaxation time of the process. Suppose that our purpose is to determine the value of $\theta$. A conventional strategy might be to perform the POVM associated with the observable $A:=\ket{0}\bra{0}$ on $\rho_\theta$. This measurement has two potential outcomes $1$ and $0$, with the probabilities $\theta$ and $1-\theta$, respectively. Intuitively, one may think of $1$ and $0$ as the head and tail of a coin, with $\theta$ corresponding to the coin's propensity to land heads. Clearly, a single POVM yields either $1$ or $0$, which is insufficient to determine $\theta$. Hence, one needs to prepare $N$ copies of $\rho_\theta$ and perform $N$ POVMs, so that a series of data, $x_1,\cdots,x_N$, with $x_i\in\{1,0\}$, are obtained. Then the value of $\theta$ can be estimated as,
$\hat{\theta}(\bm{x}):=\sum_{i=1}^Nx_i/N$, amounting to the frequency of outcomes $1$ appearing in the data $\bm{x}:=(x_1,\cdots,x_N)$. Using well-known results regarding an $N$-trial coin flip experiment, we have that the estimation error reads
\begin{eqnarray}\label{Ex-error}
\delta\theta:=\sqrt{\textrm{Var}(\hat{\theta})}=\sqrt{\theta(1-\theta)/N},
\end{eqnarray}
displaying the $1/\sqrt{N}$ scaling behavior. Here, $\textrm{Var}(\hat{\theta})$ is the variance of $\hat{\theta}(\bm{x})$. In Supplemental Material \cite{2019SM}, we show that
the above strategy is optimal among conventional strategies. That is, given the same amount of resources, one cannot improve the precision any further with conventional strategies.

Keeping this example in mind, we proceed to develop an approach going beyond conventional strategies. Suppose that we are given a dissipative system $\mathscr{S}$ with a Liouvillian superoperator $\mathcal{L}_\theta$. Here, we take $\theta$ to be a single parameter, leaving the extension to the multi-parameter case to the end of this Letter. $\mathcal{L}_\theta$ is assumed to be such that: (a) it admits a unique steady state $\rho_\theta$; (b) the nonzero eigenvalues of $\mathcal{L}_\theta$ have negative real parts, that is, there is a dissipative gap in the Liouvillian spectrum. $\mathscr{S}$ is initially prepared in its steady state $\rho_\theta$. Note that this is achievable even though $\theta$ is unknown, as $\mathscr{S}$ automatically approaches $\rho_\theta$ because of the dissipative gap.
To measure the value of an observable $A$ in the state $\rho_\theta$, we add an interaction term,
$H_I=T^{-1}A\otimes \hat{p}$,
coupling $\mathscr{S}$ to a measuring apparatus $\mathscr{A}$, with  coordinate and momentum denoted by $\hat{q}$ and $\hat{p}$, respectively.
Here, $T$ is a  positive real number, which, in the spirit of the adiabatic theorem, will be eventually sent to infinity. The dynamics of the coupling procedure is described by the following equation,
\begin{eqnarray}\label{eq:measuring}
\frac{d}{dt}\rho(t)=\mathcal{L}_\theta\rho(t)-i\left[H_I,\rho(t)\right]=:\mathcal{L}\rho(t).
\end{eqnarray}
If the coupling time is $T$ (so that the product of the coupling strength, i.e., $T^{-1}$, and the coupling time is unity), $\mathscr{S}+\mathscr{A}$ undergoes the dynamical map, $\mathcal{E}_T:=e^{\mathcal{L}T}$, transforming the initial state $\rho_\theta\otimes\ket{\phi}\bra{\phi}$ to the state $\mathcal{E}_T(\rho_\theta\otimes\ket{\phi}\bra{\phi})$ at time $T$. Here, $\ket{\phi}$ denotes the initial state of $\mathscr{A}$, which is set to be a Gaussian centered at $q=0$ with a small standard deviation, say, $1/10$ (see Fig.~\ref{fig1}). After the coupling procedure, the coordinate $\hat{q}$ is observed, in order to determine the reading of the pointer.

To figure out the effect of $\mathcal{E}_T$, we make use of the fact
$\mathcal{L}(\rho\otimes\ket{p}\bra{p^\prime})=
(\mathcal{L}_{p,p^\prime}\rho)\otimes\ket{p}\bra{p^\prime}$, with $\mathcal{L}_{p,p^\prime}\rho:=\mathcal{L}_\theta\rho-iT^{-1}(pA\rho-p^\prime \rho A)$. Here, $\ket{p}$ denotes the eigenstate of $\hat{p}$, i.e., $\hat{p}\ket{p}=p\ket{p}$. This leads to
$\mathcal{E}_T(\rho_\theta\otimes\ket{p}\bra{p^\prime})
=(e^{\mathcal{L}_{p,p^\prime}T}\rho_\theta)\otimes\ket{p}\bra{p^\prime}$.
Note that a technical result in Ref.~\cite{2014Zanardi240406} is
\begin{eqnarray}\label{Zanardi}
\norm{e^{\mathcal{L}_{p,p^\prime}T}\mathcal{P}_\theta
-e^{\widetilde{\mathcal{L}}_{p,p^\prime}T}\mathcal{P}_\theta}=O(1/T),
\end{eqnarray}
indicating that $e^{\mathcal{L}_{p,p^\prime}T}\mathcal{P}_\theta$ gets closer and closer to $e^{\widetilde{\mathcal{L}}_{p,p^\prime}T}\mathcal{P}_\theta$ as $T$ approaches infinity \cite{1note}. Here, $\widetilde{\mathcal{L}}_{p,p^\prime}:=\mathcal{P}_\theta
\mathcal{L}_{p,p^\prime}\mathcal{P}_\theta$, and $\mathcal{P}_\theta$ denotes the projection, $\mathcal{P}_\theta(X):=(\tr_\mathscr{S}X)\rho_\theta$, mapping an arbitrary operator $X$ into $\rho_\theta$. Using Eq.~(\ref{Zanardi}) and noting that the explicit expression of $\widetilde{\mathcal{L}}_{p,p^\prime}$ reads
$\widetilde{\mathcal{L}}_{p,p^\prime}=-iT^{-1}(p-p^\prime)\expt{A}_\theta
\mathcal{P}_\theta$, where $\expt{A}_\theta:=\tr(A\rho_\theta)$, we obtain
\begin{eqnarray}\label{eq:ES}
\lim_{T\rightarrow\infty}\mathcal{E}_T(\rho_\theta\otimes\ket{p}\bra{p^\prime})
=\rho_\theta\otimes e^{-i(p-p^\prime)\expt{A}_\theta}\ket{p}\bra{p^\prime}.
\end{eqnarray}
Now, expressing $\ket{\phi}$ in $\mathcal{E}_T(\rho_\theta\otimes
\ket{\phi}\bra{\phi})$ as $\ket{\phi}=\int\phi(p)\ket{p}dp$ and using the linearity of the map $\mathcal{E}_T$ as well as Eq.~(\ref{eq:ES}), we reach the first main formula of this Letter,
\begin{eqnarray}\label{1st-main-formula}
\lim_{T\rightarrow\infty}\mathcal{E}_T(\rho_\theta\otimes
\ket{\phi}\bra{\phi})=
\rho_\theta\otimes e^{-i\expt{A}_\theta\hat{p}}\ket{\phi}\bra{\phi}
e^{i\expt{A}_\theta\hat{p}}.
\end{eqnarray}
Formula (\ref{1st-main-formula})
shows that in the weak coupling and long-time limit, the steady state $\rho_\theta$ does not collapse and the pointer is shifted by the expectation value $\expt{A}_\theta$ rather than eigenvalues of $A$.

To gain physical insight into the above result, we compare our proposal of measurements (i.e., DAMs) with POVMs. Both DAMs and POVMs utilize interaction terms of the form, $H_I=g(t)A\otimes\hat{p}$, with $g(t)$ normalized to $\int g(t)dt=1$. In POVMs, this term is impulsive, that is, $g(t)$ takes an extremely large value but only for a very short time interval. Hence, the dominating term in Eq.~(\ref{eq:measuring}) is $H_I$ and the associated evolution operator reads $e^{-iA\otimes\hat{p}}$. Evidently, this operator creates
correlations between $\mathscr{S}$ and $\mathscr{A}$, giving rise to the quantum back action that the configuration of $\mathscr{S}$ after the measurement is determined by the outcome of $\mathscr{A}$.
Contrary to POVMs, DAMs exploit the opposite limit of an extremely weak but long-time interaction, i.e., $g(t)=T^{-1}$. For this, the dissipative term $\mathcal{L}_\theta$ dominates the interaction term $H_I$ in Eq.~(\ref{eq:measuring}). The former effectively eliminates correlations created by the latter through continuously projecting $\mathscr{S}$ into its steady state $\rho_\theta$. Resulted from this nontrivial interplay of the two terms is the decoupling of $\mathscr{S}$ and $\mathscr{A}$ in the long-time limit, which inhibits the state of $\mathscr{S}$ from any change or collapse. Such a mechanism is suggestive of the quantum Zeno effect, making DAMs distinct  from POVMs in nature.

Ideally, the evolved state of $\mathscr{A}$ (at time $T$) is the Gaussian centered at $q=\expt{A}_\theta$ (see Fig.~\ref{fig1}b). However, as $T$ is finite in practice, there exist non-adiabatic effects, leading to the consequence that correlations between $\mathscr{S}$ and $\mathscr{A}$ are not completely eliminated. This may cause slight deviations of the evolved state from the ideal Gaussian (see Fig.~\ref{fig1}c). To quantify such deviations, we may use the following measure,
\begin{eqnarray}\label{non-adiabaticity}
\Delta:=\norm{
\tr_\mathscr{S}\mathcal{E}_T(\rho_\theta\otimes\ket{\phi}\bra{\phi})
-\tr_\mathscr{S}\lim_{T\rightarrow\infty}\mathcal{E}_T
(\rho_\theta\otimes\ket{\phi}\bra{\phi})}.\nonumber
\end{eqnarray}
Noting that $\mathcal{E}_T(\rho_\theta\otimes\ket{p}\bra{p^\prime})
=(e^{\mathcal{L}_{p,p^\prime}T}\rho_\theta)\otimes\ket{p}\bra{p^\prime}$ and $\lim_{T\rightarrow\infty}\mathcal{E}_T(\rho_\theta\otimes\ket{p}\bra{p^\prime})
=(e^{\widetilde{\mathcal{L}}_{p,p^\prime}T}\rho_\theta)\otimes\ket{p}
\bra{p^\prime}$, we can rewrite $\Delta$ as
\begin{eqnarray}\label{ins-delta}
\Delta=\left(\iint dp dp^\prime\abs{\phi(p)}^2\abs{\phi(p^\prime)}^2
\abs{\Delta({p,p^\prime})}^2\right)^{1/2},
\end{eqnarray}
where $\Delta({p,p^\prime}):=\tr_\mathscr{S}(e^{\mathcal{L}_{p,p^\prime}T}
\rho_\theta-e^{\widetilde{\mathcal{L}}_{p,p^\prime}T}
\rho_\theta)$. Using Eq.~(\ref{Zanardi}) as well as the Cauchy-Schwarz inequality $\abs{\tr X^\dagger Y}\leq\norm{X}\norm{Y}$,
we have
$\abs{\Delta({p,p^\prime})}=O(1/T)$. Substituting this equality into Eq.~(\ref{ins-delta}) gives
\begin{eqnarray}\label{non-adiabatic-effects}
\Delta=O(1/T),
\end{eqnarray}
namely, the deviations of the evolved state from the ideal Gaussian are smaller than $cT^{-1}$. The prefactor $c$ is related to the dissipative gap. Roughly speaking, $c$ and, therefore, the deviations decrease if the dissipative gap increases.

\begin{figure}[htbp]
\includegraphics[width=0.45\textwidth]{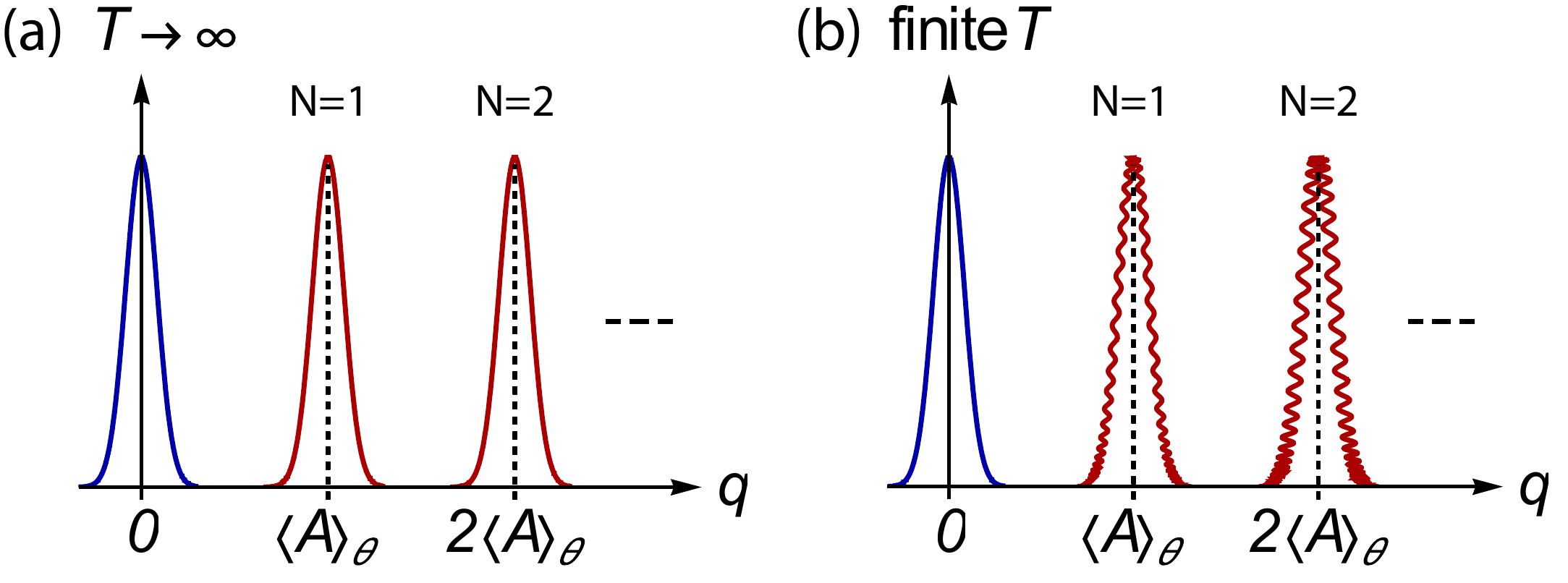}
\caption{Schematic representation of the evolved state of the apparatus with coupling strength $T^{-1}$ and coupling time $NT$.}
\label{fig2}
\end{figure}

Having proposed DAMs, we now use them to solve the estimation problem. To make our idea clear, let us leave alone non-adiabatic effects for now. We observe that the expectation value $\expt{A}_\theta$ is directly related to the parameter $\theta$. Indeed, it is not difficult to choose an observable $A$ such that the function $f:\theta\mapsto\expt{A}_\theta$ is invertible. Then there exists an inverse function $f^{-1}$ directly extracting the value of $\theta$ from $\expt{A}_\theta$. Note that the pointer reading in the DAM measuring $A$ is a random variable $q$ that takes values equal to or close to $\expt{A}_\theta$, as it fulfills the Gaussian distribution with mean value $\expt{A}_\theta$ (see Fig.~\ref{fig1}b). Applying $f^{-1}$ to $q$ gives an estimation of $\theta$; that is, $\hat{\theta}(q)=f^{-1}(q)$. This is an unbiased estimator, for which the estimation error reads $\delta\theta=\sigma/\abs{\frac{\partial f}{\partial\theta}}$, where
$\sigma$ denotes the standard deviation of the Gaussian distribution. To improve the precision, we need to employ more resources in the estimation procedure. In contrast to conventional strategies, where employing more resources (often) means repeating the measurement $N$ times, we here advocate the use of another kind of resources, i.e., increasing the coupling time from $T$ to $NT$ (while maintaining the coupling strength at $T^{-1}$).  For such a scenario, the dynamical map describing the coupling procedure is $e^{\mathcal{L}NT}=\mathcal{E}_T^N$. Hence, the pointer is shifted by $N\expt{A}_\theta$ (see Fig.~\ref{fig2}a),
indicating that its reading $q$ takes values equal to or close to $N\expt{A}_\theta$ now. Accordingly, the estimator should be changed as $\hat{\theta}(q)=f^{-1}(q/N)$. As $q/N$ is a random variable with standard deviation $\sigma/N$, the estimation error is improved to $\delta\theta=\sigma/(N\abs{\frac{\partial f}{\partial\theta}})$, i.e.,  the $1/N$ scaling of precision.

In light of the above analysis, we may now specify our approach as follows:
(i) Prepare $\mathscr{S}$ in $\rho_\theta$ and choose an observable $A$ such that $f:\theta\mapsto\expt{A}_\theta$ is invertible. (ii) Measure $A$ by performing the associated DAM with coupling time $NT$. (iii) Estimate the value of $\theta$ from the pointer reading $q$ as $\hat{\theta}(q)=f^{-1}(q/N)$. To fully understand the estimation error of our approach, we need to take into account non-adiabatic effects. Such effects cause deviations of the probability distribution of $q$ from the ideal Gaussian, thereby introducing an additional error in our approach. Intuitively speaking, this additional error decreases as $T$ increases, as indicated by Eq.~(\ref{non-adiabatic-effects}). Nevertheless, it may be an increasing function of $N$, as the deviations may get
larger and larger as $N$ increases (see Fig.~\ref{fig2}b). Indeed, detailed analyses \cite{2019SM} show that the estimation error with the additional error included  reads
\begin{eqnarray}\label{2st-main-formula}
\delta\theta=\sqrt{\sigma^2+O(N/T)+O(N^2/T^2)}
/\left(N\abs{\frac{\partial f}{\partial\theta}}\right),
\end{eqnarray}
with $O(N/T)=-2 N\textrm{Re}\left(
\tr_\mathscr{S}\left[A\mathcal{S}_\theta(A\rho_\theta)\right]\right)/{T}$ and
$O(N^2/T^2)=N^2\textrm{Im}\left(\tr_\mathscr{S}
[A\mathcal{S}_\theta(A\rho_\theta)]\right)^2
/(T^2\sigma^2)$.
Here, $\mathcal{S}_\theta:=-\int_0^\infty dt e^{t\mathcal{L}_\theta}\mathcal{Q}_\theta$ is the pseudoinverse of $\mathcal{L}_\theta$, i.e., $\mathcal{L}_\theta\mathcal{S}_\theta=
\mathcal{S}_\theta\mathcal{L}_\theta=\mathcal{Q}_\theta$, where $\mathcal{Q}_\theta:=1-\mathcal{P}_\theta$.
Equation (\ref{2st-main-formula}) is the second main formula of this Letter. As can be seen from Eq.~(\ref{2st-main-formula}), so long as $N\leq N_\textrm{max}:=O(T)$, the numerator on the RHS of Eq.~(\ref{2st-main-formula}) is of the order $O(1)$, indicating that $\delta\theta\thicksim 1/N$, that is, our approach gives the Heisenberg scaling of precision. Although this pleasant scaling saturates for a given $T$, the saturation point $N_\textrm{max}$ keeps increasing as $T$ increases, thus there is no fundamental limit. Furthermore, given a concrete model of $\mathscr{S}$, we can obtain the explicit expression of $\mathcal{S}_\theta$ and hence that of Eq.~(\ref{2st-main-formula}), based on which we can choose proper $T$ and $N$ to achieve a desired precision.

To illustrate the application of our approach, consider again the foregoing example. For the purpose of comparing our approach with the conventional strategy, we choose the same observable $A=\ket{0}\bra{0}$, so that $\expt{A}_\theta=\theta$, $f:\theta\mapsto\theta$, and $f^{-1}=f$. Then, the outcome of the DAM with coupling time $NT$ is a random variable $q$ taking values equal to or close to $N\theta$. This is in sharp contrast to the very nature of the associated POVM, whose outcome is a random variable taking values $1$ and $0$, which is not directly related to $\theta$. Thanks to this, unlike the conventional strategy, which needs to perform many measurements such that sufficient data are obtained and $\theta$ can be inferred indirectly, our approach directly extracts the value of $\theta$ from the single measurement result $q$ as $\hat{\theta}(q)=q/N$. Noting that $\mathcal{S}_\theta(X)=\mathcal{P}_\theta(X)-X-\ket{0}\bra{0}X\ket{1}\bra{1}-
\ket{1}\bra{1}X\ket{0}\bra{0}$, we deduce from Eq.~(\ref{2st-main-formula}) that the estimation error reads
\begin{eqnarray}\label{Ex-error-our}
\delta\theta=\sqrt{\sigma^2+\frac{2\theta(1-\theta)N}{T}}/N,
\end{eqnarray}
displaying the $1/N$ scaling behavior as long as $N\leq N_\textrm{max}=O(T)$. Therefore, compared with the conventional strategy, our approach provides a quadratic improvement on precision, but at the cost of a single measurement.

Let us end the development of our approach by addressing the multi-parameter case. Suppose that there are $M$ parameters, namely, $\bm{\theta}=(\theta_1,\cdots,\theta_M)$.
For this, we need to perform $M$ DAMs, each of which is associated with an observable $A_i$, $i\in\{1,\cdots,M\}$. Here, the observables are chosen
such that the function $\bm{f}:(\theta_1,\cdots,\theta_M)\mapsto (\expt{A_1}_\theta,\cdots,\expt{A_M}_\theta)$ is invertible. Then $\bm{f}^{-1}$ exists and has the effect $\bm{f}^{-1}(\expt{A_1}_\theta,\cdots,\expt{A_M}_\theta)= (\theta_1,\cdots,\theta_M)$. Note that the pointer reading of the DAM associated with $A_i$ is a random variable $q_i$ taking values equal to or close to $N\expt{A_i}_\theta$, provided that the measured state is $\rho_\theta$ and the coupling time is $NT$. We can estimate the values of these parameters as
$\hat{\bm{\theta}}(\bm{q})=\bm{f}^{-1}(\bm{q}/N)$, where $\bm{q}:=(q_1,\cdots,q_M)$. As long as $N\leq N_\textrm{max}=O(T)$, there is $\delta\bm{\theta}\thicksim1/N$ \cite{2019SM}, indicating that our approach gives the Heisenberg scaling of precision in the multi-parameter case as well. The formula describing the error $\delta\bm{\theta}$ is a simple generalization of Eq.~(\ref{2st-main-formula})
and is given in Supplemental Material \cite{2019SM}.

Before concluding, we point out that DAMs are distinct from AAMs
in a number of aspects. First, the measured system in an AAM is a closed system with an energy gap, whereas that in a DAM is an open system with a dissipative gap. Second, the outcomes of AAMs are restricted to expectation values in pure states, whereas those of DAMs can be expectation values in any mixed states. Third, decoherence and dissipation are detrimental to AAMs, whereas they play a positive role in DAMs. Besides, it is worth noting that AAMs have been experimentally realized in an optical setup \cite{2017Piacentini1191}, wherein decoherence and dissipation are suppressed via an active protection technique. Instead of actively suppressing decoherence and dissipation, one may passively exploit them to realize DAMs in this setup.

In conclusion, we have developed a direct and efficient approach to estimation of parameters characterizing dissipative processes. The key is to propose a new kind of measurements, featured by an extremely weak but long-time interaction between the dissipative system in question and its measuring apparatus. The dissipative dynamics of the system continuously eliminates correlations created by the weak interaction, resulting in the decoupling of the system from the measuring apparatus in the long-time limit. Unlike POVMs, our measurement therefore does not collapse the state to be measured, and moreover, its outcome is the expectation value of an observable, which is directly related to the parameters of interest. By virtue of this, our approach is able to extract the values of the parameters in a straightforward way, essentially offering the Heisenberg scaling of precision, as long as the non-adiabaticity-induced error is kept small. We highlight that our approach works in a state-protective fashion with only $M$ measurements for $M$ parameters to be estimated. These experiment-friendly features should be immensely useful in the context of dissipative quantum
information processing. Our findings, solidified by a simple yet well-known example, provides a fundamentally new route to solve quantum estimation problems.

\begin{acknowledgments}
J.G.~is supported  by the
Singapore NRF Grant No.~NRF-NRFI2017-04 (WBS No.~R-144-000-378-281).
D.-J.Z.~acknowledges support from the National Natural Science Foundation of
China through Grant No.~11705105 before he joined NUS.
\end{acknowledgments}



\begin{thebibliography}{34}%
\makeatletter
\providecommand \@ifxundefined [1]{%
 \@ifx{#1\undefined}
}%
\providecommand \@ifnum [1]{%
 \ifnum #1\expandafter \@firstoftwo
 \else \expandafter \@secondoftwo
 \fi
}%
\providecommand \@ifx [1]{%
 \ifx #1\expandafter \@firstoftwo
 \else \expandafter \@secondoftwo
 \fi
}%
\providecommand \natexlab [1]{#1}%
\providecommand \enquote  [1]{``#1''}%
\providecommand \bibnamefont  [1]{#1}%
\providecommand \bibfnamefont [1]{#1}%
\providecommand \citenamefont [1]{#1}%
\providecommand \href@noop [0]{\@secondoftwo}%
\providecommand \href [0]{\begingroup \@sanitize@url \@href}%
\providecommand \@href[1]{\@@startlink{#1}\@@href}%
\providecommand \@@href[1]{\endgroup#1\@@endlink}%
\providecommand \@sanitize@url [0]{\catcode `\\12\catcode `\$12\catcode
  `\&12\catcode `\#12\catcode `\^12\catcode `\_12\catcode `\%12\relax}%
\providecommand \@@startlink[1]{}%
\providecommand \@@endlink[0]{}%
\providecommand \url  [0]{\begingroup\@sanitize@url \@url }%
\providecommand \@url [1]{\endgroup\@href {#1}{\urlprefix }}%
\providecommand \urlprefix  [0]{URL }%
\providecommand \Eprint [0]{\href }%
\providecommand \doibase [0]{http://dx.doi.org/}%
\providecommand \selectlanguage [0]{\@gobble}%
\providecommand \bibinfo  [0]{\@secondoftwo}%
\providecommand \bibfield  [0]{\@secondoftwo}%
\providecommand \translation [1]{[#1]}%
\providecommand \BibitemOpen [0]{}%
\providecommand \bibitemStop [0]{}%
\providecommand \bibitemNoStop [0]{.\EOS\space}%
\providecommand \EOS [0]{\spacefactor3000\relax}%
\providecommand \BibitemShut  [1]{\csname bibitem#1\endcsname}%
\let\auto@bib@innerbib\@empty
\bibitem [{\citenamefont {Paris}(2009)}]{2009Paris125}%
  \BibitemOpen
  \bibfield  {author} {\bibinfo {author} {\bibfnamefont {M.~G.~A.}\
  \bibnamefont {Paris}},\ }\bibfield  {title} {\enquote {\bibinfo {title}
  {Quantum estimation for quantum technology},}\ }\href {\doibase
  10.1142/S0219749909004839} {\bibfield  {journal} {\bibinfo  {journal} {Int.
  J. Quantum Inform.}\ }\textbf {\bibinfo {volume} {07}},\ \bibinfo {pages}
  {125} (\bibinfo {year} {2009})}\BibitemShut {NoStop}%
\bibitem [{\citenamefont {Giovannetti}\ \emph {et~al.}(2006)\citenamefont
  {Giovannetti}, \citenamefont {Lloyd},\ and\ \citenamefont
  {Maccone}}]{2006Giovannetti10401}%
  \BibitemOpen
  \bibfield  {author} {\bibinfo {author} {\bibfnamefont {V.}~\bibnamefont
  {Giovannetti}}, \bibinfo {author} {\bibfnamefont {S.}~\bibnamefont {Lloyd}},
  \ and\ \bibinfo {author} {\bibfnamefont {L.}~\bibnamefont {Maccone}},\
  }\bibfield  {title} {\enquote {\bibinfo {title} {Quantum metrology},}\ }\href
  {\doibase 10.1103/PhysRevLett.96.010401} {\bibfield  {journal} {\bibinfo
  {journal} {Phys. Rev. Lett.}\ }\textbf {\bibinfo {volume} {96}},\ \bibinfo
  {pages} {010401} (\bibinfo {year} {2006})}\BibitemShut {NoStop}%
\bibitem [{\citenamefont {Braunstein}\ and\ \citenamefont
  {Caves}(1994)}]{1994Braunstein3439}%
  \BibitemOpen
  \bibfield  {author} {\bibinfo {author} {\bibfnamefont {S.~L.}\ \bibnamefont
  {Braunstein}}\ and\ \bibinfo {author} {\bibfnamefont {C.~M.}\ \bibnamefont
  {Caves}},\ }\bibfield  {title} {\enquote {\bibinfo {title} {Statistical
  distance and the geometry of quantum states},}\ }\href {\doibase
  10.1103/PhysRevLett.72.3439} {\bibfield  {journal} {\bibinfo  {journal}
  {Phys. Rev. Lett.}\ }\textbf {\bibinfo {volume} {72}},\ \bibinfo {pages}
  {3439} (\bibinfo {year} {1994})}\BibitemShut {NoStop}%
\bibitem [{\citenamefont {Giovannetti}\ \emph {et~al.}(2004)\citenamefont
  {Giovannetti}, \citenamefont {Lloyd},\ and\ \citenamefont
  {Maccone}}]{2004Giovannetti1330}%
  \BibitemOpen
  \bibfield  {author} {\bibinfo {author} {\bibfnamefont {V.}~\bibnamefont
  {Giovannetti}}, \bibinfo {author} {\bibfnamefont {S.}~\bibnamefont {Lloyd}},
  \ and\ \bibinfo {author} {\bibfnamefont {L.}~\bibnamefont {Maccone}},\
  }\bibfield  {title} {\enquote {\bibinfo {title} {Quantum-enhanced
  measurements: Beating the standard quantum limit},}\ }\href {\doibase
  10.1126/science.1104149} {\bibfield  {journal} {\bibinfo  {journal}
  {Science}\ }\textbf {\bibinfo {volume} {306}},\ \bibinfo {pages} {1330}
  (\bibinfo {year} {2004})}\BibitemShut {NoStop}%
\bibitem [{\citenamefont {Giovannetti}\ \emph {et~al.}(2011)\citenamefont
  {Giovannetti}, \citenamefont {Lloyd},\ and\ \citenamefont
  {Maccone}}]{2011Giovannetti222}%
  \BibitemOpen
  \bibfield  {author} {\bibinfo {author} {\bibfnamefont {V.}~\bibnamefont
  {Giovannetti}}, \bibinfo {author} {\bibfnamefont {S.}~\bibnamefont {Lloyd}},
  \ and\ \bibinfo {author} {\bibfnamefont {L.}~\bibnamefont {Maccone}},\
  }\bibfield  {title} {\enquote {\bibinfo {title} {Advances in quantum
  metrology},}\ }\href {\doibase 10.1038/nphoton.2011.35} {\bibfield  {journal}
  {\bibinfo  {journal} {Nat. Photonics}\ }\textbf {\bibinfo {volume} {5}},\
  \bibinfo {pages} {222} (\bibinfo {year} {2011})}\BibitemShut {NoStop}%
\bibitem [{\citenamefont {Huelga}\ \emph {et~al.}(1997)\citenamefont {Huelga},
  \citenamefont {Macchiavello}, \citenamefont {Pellizzari}, \citenamefont
  {Ekert}, \citenamefont {Plenio},\ and\ \citenamefont
  {Cirac}}]{1997Huelga3865}%
  \BibitemOpen
  \bibfield  {author} {\bibinfo {author} {\bibfnamefont {S.~F.}\ \bibnamefont
  {Huelga}}, \bibinfo {author} {\bibfnamefont {C.}~\bibnamefont
  {Macchiavello}}, \bibinfo {author} {\bibfnamefont {T.}~\bibnamefont
  {Pellizzari}}, \bibinfo {author} {\bibfnamefont {A.~K.}\ \bibnamefont
  {Ekert}}, \bibinfo {author} {\bibfnamefont {M.~B.}\ \bibnamefont {Plenio}}, \
  and\ \bibinfo {author} {\bibfnamefont {J.~I.}\ \bibnamefont {Cirac}},\
  }\bibfield  {title} {\enquote {\bibinfo {title} {Improvement of frequency
  standards with quantum entanglement},}\ }\href {\doibase
  10.1103/PhysRevLett.79.3865} {\bibfield  {journal} {\bibinfo  {journal}
  {Phys. Rev. Lett.}\ }\textbf {\bibinfo {volume} {79}},\ \bibinfo {pages}
  {3865} (\bibinfo {year} {1997})}\BibitemShut {NoStop}%
\bibitem [{\citenamefont {Monras}\ and\ \citenamefont
  {Paris}(2007)}]{2007Monras160401}%
  \BibitemOpen
  \bibfield  {author} {\bibinfo {author} {\bibfnamefont {A.}~\bibnamefont
  {Monras}}\ and\ \bibinfo {author} {\bibfnamefont {M.~G.~A.}\ \bibnamefont
  {Paris}},\ }\bibfield  {title} {\enquote {\bibinfo {title} {Optimal quantum
  estimation of loss in bosonic channels},}\ }\href {\doibase
  10.1103/PhysRevLett.98.160401} {\bibfield  {journal} {\bibinfo  {journal}
  {Phys. Rev. Lett.}\ }\textbf {\bibinfo {volume} {98}},\ \bibinfo {pages}
  {160401} (\bibinfo {year} {2007})}\BibitemShut {NoStop}%
\bibitem [{\citenamefont {Dorner}\ \emph {et~al.}(2009)\citenamefont {Dorner},
  \citenamefont {Demkowicz-Dobrzanski}, \citenamefont {Smith}, \citenamefont
  {Lundeen}, \citenamefont {Wasilewski}, \citenamefont {Banaszek},\ and\
  \citenamefont {Walmsley}}]{2009Dorner40403a}%
  \BibitemOpen
  \bibfield  {author} {\bibinfo {author} {\bibfnamefont {U.}~\bibnamefont
  {Dorner}}, \bibinfo {author} {\bibfnamefont {R.}~\bibnamefont
  {Demkowicz-Dobrzanski}}, \bibinfo {author} {\bibfnamefont {B.~J.}\
  \bibnamefont {Smith}}, \bibinfo {author} {\bibfnamefont {J.~S.}\ \bibnamefont
  {Lundeen}}, \bibinfo {author} {\bibfnamefont {W.}~\bibnamefont {Wasilewski}},
  \bibinfo {author} {\bibfnamefont {K.}~\bibnamefont {Banaszek}}, \ and\
  \bibinfo {author} {\bibfnamefont {I.~A.}\ \bibnamefont {Walmsley}},\
  }\bibfield  {title} {\enquote {\bibinfo {title} {Optimal quantum phase
  estimation},}\ }\href {\doibase 10.1103/PhysRevLett.102.040403} {\bibfield
  {journal} {\bibinfo  {journal} {Phys. Rev. Lett.}\ }\textbf {\bibinfo
  {volume} {102}},\ \bibinfo {pages} {040403} (\bibinfo {year}
  {2009})}\BibitemShut {NoStop}%
\bibitem [{\citenamefont {Escher}\ \emph {et~al.}(2011)\citenamefont {Escher},
  \citenamefont {de~Matos~Filho},\ and\ \citenamefont
  {Davidovich}}]{2011Escher406}%
  \BibitemOpen
  \bibfield  {author} {\bibinfo {author} {\bibfnamefont {B.~M.}\ \bibnamefont
  {Escher}}, \bibinfo {author} {\bibfnamefont {R.~L.}\ \bibnamefont
  {de~Matos~Filho}}, \ and\ \bibinfo {author} {\bibfnamefont {L.}~\bibnamefont
  {Davidovich}},\ }\bibfield  {title} {\enquote {\bibinfo {title} {General
  framework for estimating the ultimate precision limit in noisy
  quantum-enhanced metrology},}\ }\href {\doibase 10.1038/NPHYS1958} {\bibfield
   {journal} {\bibinfo  {journal} {Nat. Phys.}\ }\textbf {\bibinfo {volume}
  {7}},\ \bibinfo {pages} {406} (\bibinfo {year} {2011})}\BibitemShut {NoStop}%
\bibitem [{\citenamefont {Demkowicz-Dobrza\'{n}ski}\ \emph
  {et~al.}(2012)\citenamefont {Demkowicz-Dobrza\'{n}ski}, \citenamefont
  {Ko{\l}ody\'{n}ski},\ and\ \citenamefont
  {Gu\c{t}\u{a}}}]{2012Demkowicz-Dobrzanski1063}%
  \BibitemOpen
  \bibfield  {author} {\bibinfo {author} {\bibfnamefont {R.}~\bibnamefont
  {Demkowicz-Dobrza\'{n}ski}}, \bibinfo {author} {\bibfnamefont
  {J.}~\bibnamefont {Ko{\l}ody\'{n}ski}}, \ and\ \bibinfo {author}
  {\bibfnamefont {M.}~\bibnamefont {Gu\c{t}\u{a}}},\ }\bibfield  {title}
  {\enquote {\bibinfo {title} {The elusive Heisenberg limit in quantum-enhanced
  metrology},}\ }\href {\doibase 10.1038/ncomms2067} {\bibfield  {journal}
  {\bibinfo  {journal} {Nat. Commun.}\ }\textbf {\bibinfo {volume} {3}},\
  \bibinfo {pages} {1063} (\bibinfo {year} {2012})}\BibitemShut {NoStop}%
\bibitem [{\citenamefont {Alipour}\ \emph {et~al.}(2014)\citenamefont
  {Alipour}, \citenamefont {Mehboudi},\ and\ \citenamefont
  {Rezakhani}}]{2014Alipour120405}%
  \BibitemOpen
  \bibfield  {author} {\bibinfo {author} {\bibfnamefont {S.}~\bibnamefont
  {Alipour}}, \bibinfo {author} {\bibfnamefont {M.}~\bibnamefont {Mehboudi}}, \
  and\ \bibinfo {author} {\bibfnamefont {A.~T.}\ \bibnamefont {Rezakhani}},\
  }\bibfield  {title} {\enquote {\bibinfo {title} {Quantum metrology in open
  systems: Dissipative cram\'{e}r-rao bound},}\ }\href {\doibase
  10.1103/PhysRevLett.112.120405} {\bibfield  {journal} {\bibinfo  {journal}
  {Phys. Rev. Lett.}\ }\textbf {\bibinfo {volume} {112}},\ \bibinfo {pages}
  {120405} (\bibinfo {year} {2014})}\BibitemShut {NoStop}%
\bibitem [{\citenamefont {Diehl}\ \emph {et~al.}(2008)\citenamefont {Diehl},
  \citenamefont {Micheli}, \citenamefont {Kantian}, \citenamefont {Kraus},
  \citenamefont {B\"{u}chler},\ and\ \citenamefont {Zoller}}]{2008Diehl878}%
  \BibitemOpen
  \bibfield  {author} {\bibinfo {author} {\bibfnamefont {S.}~\bibnamefont
  {Diehl}}, \bibinfo {author} {\bibfnamefont {A.}~\bibnamefont {Micheli}},
  \bibinfo {author} {\bibfnamefont {A.}~\bibnamefont {Kantian}}, \bibinfo
  {author} {\bibfnamefont {B.}~\bibnamefont {Kraus}}, \bibinfo {author}
  {\bibfnamefont {H.~P.}\ \bibnamefont {B\"{u}chler}}, \ and\ \bibinfo {author}
  {\bibfnamefont {P.}~\bibnamefont {Zoller}},\ }\bibfield  {title} {\enquote
  {\bibinfo {title} {Quantum states and phases in driven open quantum systems
  with cold atoms},}\ }\href {\doibase 10.1038/nphys1073} {\bibfield  {journal}
  {\bibinfo  {journal} {Nat. Phys.}\ }\textbf {\bibinfo {volume} {4}},\
  \bibinfo {pages} {878} (\bibinfo {year} {2008})}\BibitemShut {NoStop}%
\bibitem [{\citenamefont {Kastoryano}\ \emph {et~al.}(2011)\citenamefont
  {Kastoryano}, \citenamefont {Reiter},\ and\ \citenamefont
  {S{\o}rensen}}]{2011Kastoryano90502}%
  \BibitemOpen
  \bibfield  {author} {\bibinfo {author} {\bibfnamefont {M.~J.}\ \bibnamefont
  {Kastoryano}}, \bibinfo {author} {\bibfnamefont {F.}~\bibnamefont {Reiter}},
  \ and\ \bibinfo {author} {\bibfnamefont {A.~S.}\ \bibnamefont
  {S{\o}rensen}},\ }\bibfield  {title} {\enquote {\bibinfo {title} {Dissipative
  preparation of entanglement in optical cavities},}\ }\href {\doibase
  10.1103/PhysRevLett.106.090502} {\bibfield  {journal} {\bibinfo  {journal}
  {Phys. Rev. Lett.}\ }\textbf {\bibinfo {volume} {106}},\ \bibinfo {pages}
  {090502} (\bibinfo {year} {2011})}\BibitemShut {NoStop}%
\bibitem [{\citenamefont {Cho}\ \emph {et~al.}(2011)\citenamefont {Cho},
  \citenamefont {Bose},\ and\ \citenamefont {Kim}}]{2011Cho20504}%
  \BibitemOpen
  \bibfield  {author} {\bibinfo {author} {\bibfnamefont {J.}~\bibnamefont
  {Cho}}, \bibinfo {author} {\bibfnamefont {S.}~\bibnamefont {Bose}}, \ and\
  \bibinfo {author} {\bibfnamefont {M.~S.}\ \bibnamefont {Kim}},\ }\bibfield
  {title} {\enquote {\bibinfo {title} {Optical pumping into many-body
  entanglement},}\ }\href {\doibase 10.1103/PhysRevLett.106.020504} {\bibfield
  {journal} {\bibinfo  {journal} {Phys. Rev. Lett.}\ }\textbf {\bibinfo
  {volume} {106}},\ \bibinfo {pages} {020504} (\bibinfo {year}
  {2011})}\BibitemShut {NoStop}%
\bibitem [{\citenamefont {Krauter}\ \emph {et~al.}(2011)\citenamefont
  {Krauter}, \citenamefont {Muschik}, \citenamefont {Jensen}, \citenamefont
  {Wasilewski}, \citenamefont {Petersen}, \citenamefont {Cirac},\ and\
  \citenamefont {Polzik}}]{2011Krauter80503}%
  \BibitemOpen
  \bibfield  {author} {\bibinfo {author} {\bibfnamefont {H.}~\bibnamefont
  {Krauter}}, \bibinfo {author} {\bibfnamefont {C.~A.}\ \bibnamefont
  {Muschik}}, \bibinfo {author} {\bibfnamefont {K.}~\bibnamefont {Jensen}},
  \bibinfo {author} {\bibfnamefont {W.}~\bibnamefont {Wasilewski}}, \bibinfo
  {author} {\bibfnamefont {J.~M.}\ \bibnamefont {Petersen}}, \bibinfo {author}
  {\bibfnamefont {J.~I.}\ \bibnamefont {Cirac}}, \ and\ \bibinfo {author}
  {\bibfnamefont {E.~S.}\ \bibnamefont {Polzik}},\ }\bibfield  {title}
  {\enquote {\bibinfo {title} {Entanglement generated by dissipation and steady
  state entanglement of two macroscopic objects},}\ }\href {\doibase
  10.1103/PhysRevLett.107.080503} {\bibfield  {journal} {\bibinfo  {journal}
  {Phys. Rev. Lett.}\ }\textbf {\bibinfo {volume} {107}},\ \bibinfo {pages}
  {080503} (\bibinfo {year} {2011})}\BibitemShut {NoStop}%
\bibitem [{\citenamefont {Vollbrecht}\ \emph {et~al.}(2011)\citenamefont
  {Vollbrecht}, \citenamefont {Muschik},\ and\ \citenamefont
  {Cirac}}]{2011Vollbrecht120502}%
  \BibitemOpen
  \bibfield  {author} {\bibinfo {author} {\bibfnamefont {K.~G.~H.}\
  \bibnamefont {Vollbrecht}}, \bibinfo {author} {\bibfnamefont {C.~A.}\
  \bibnamefont {Muschik}}, \ and\ \bibinfo {author} {\bibfnamefont {J.~I.}\
  \bibnamefont {Cirac}},\ }\bibfield  {title} {\enquote {\bibinfo {title}
  {Entanglement distillation by dissipation and continuous quantum
  repeaters},}\ }\href {\doibase 10.1103/PhysRevLett.107.120502} {\bibfield
  {journal} {\bibinfo  {journal} {Phys. Rev. Lett.}\ }\textbf {\bibinfo
  {volume} {107}},\ \bibinfo {pages} {120502} (\bibinfo {year}
  {2011})}\BibitemShut {NoStop}%
\bibitem [{\citenamefont {Carr}\ and\ \citenamefont
  {Saffman}(2013)}]{2013Carr33607}%
  \BibitemOpen
  \bibfield  {author} {\bibinfo {author} {\bibfnamefont {A.~W.}\ \bibnamefont
  {Carr}}\ and\ \bibinfo {author} {\bibfnamefont {M.}~\bibnamefont {Saffman}},\
  }\bibfield  {title} {\enquote {\bibinfo {title} {Preparation of entangled and
  antiferromagnetic states by dissipative rydberg pumping},}\ }\href {\doibase
  10.1103/PhysRevLett.111.033607} {\bibfield  {journal} {\bibinfo  {journal}
  {Phys. Rev. Lett.}\ }\textbf {\bibinfo {volume} {111}},\ \bibinfo {pages}
  {033607} (\bibinfo {year} {2013})}\BibitemShut {NoStop}%
\bibitem [{\citenamefont {Torre}\ \emph {et~al.}(2013)\citenamefont {Torre},
  \citenamefont {Otterbach}, \citenamefont {Demler}, \citenamefont {Vuletic},\
  and\ \citenamefont {Lukin}}]{2013Torre120402}%
  \BibitemOpen
  \bibfield  {author} {\bibinfo {author} {\bibfnamefont {E.~G.~Dalla}\
  \bibnamefont {Torre}}, \bibinfo {author} {\bibfnamefont {J.}~\bibnamefont
  {Otterbach}}, \bibinfo {author} {\bibfnamefont {E.}~\bibnamefont {Demler}},
  \bibinfo {author} {\bibfnamefont {V.}~\bibnamefont {Vuletic}}, \ and\
  \bibinfo {author} {\bibfnamefont {M.~D.}\ \bibnamefont {Lukin}},\ }\bibfield
  {title} {\enquote {\bibinfo {title} {Dissipative preparation of spin squeezed
  atomic ensembles in a steady state},}\ }\href {\doibase
  10.1103/PhysRevLett.110.120402} {\bibfield  {journal} {\bibinfo  {journal}
  {Phys. Rev. Lett.}\ }\textbf {\bibinfo {volume} {110}},\ \bibinfo {pages}
  {120402} (\bibinfo {year} {2013})}\BibitemShut {NoStop}%
\bibitem [{\citenamefont {Rao}\ and\ \citenamefont
  {M{\o}lmer}(2013)}]{2013Rao33606}%
  \BibitemOpen
  \bibfield  {author} {\bibinfo {author} {\bibfnamefont {D.~D.~B.}\
  \bibnamefont {Rao}}\ and\ \bibinfo {author} {\bibfnamefont {K.}~\bibnamefont
  {M{\o}lmer}},\ }\bibfield  {title} {\enquote {\bibinfo {title} {Dark
  entangled steady states of interacting rydberg atoms},}\ }\href {\doibase
  10.1103/PhysRevLett.111.033606} {\bibfield  {journal} {\bibinfo  {journal}
  {Phys. Rev. Lett.}\ }\textbf {\bibinfo {volume} {111}},\ \bibinfo {pages}
  {033606} (\bibinfo {year} {2013})}\BibitemShut {NoStop}%
\bibitem [{\citenamefont {Bentley}\ \emph {et~al.}(2014)\citenamefont
  {Bentley}, \citenamefont {Carvalho}, \citenamefont {Kielpinski},\ and\
  \citenamefont {Hope}}]{2014Bentley40501}%
  \BibitemOpen
  \bibfield  {author} {\bibinfo {author} {\bibfnamefont {C.~D.~B.}\
  \bibnamefont {Bentley}}, \bibinfo {author} {\bibfnamefont {A.~R.~R.}\
  \bibnamefont {Carvalho}}, \bibinfo {author} {\bibfnamefont {D.}~\bibnamefont
  {Kielpinski}}, \ and\ \bibinfo {author} {\bibfnamefont {J.~J.}\ \bibnamefont
  {Hope}},\ }\bibfield  {title} {\enquote {\bibinfo {title} {Detection-enhanced
  steady state entanglement with ions},}\ }\href {\doibase
  10.1103/PhysRevLett.113.040501} {\bibfield  {journal} {\bibinfo  {journal}
  {Phys. Rev. Lett.}\ }\textbf {\bibinfo {volume} {113}},\ \bibinfo {pages}
  {040501} (\bibinfo {year} {2014})}\BibitemShut {NoStop}%
\bibitem [{\citenamefont {Abdi}\ \emph {et~al.}(2016)\citenamefont {Abdi},
  \citenamefont {Degenfeld-Schonburg}, \citenamefont {Sameti}, \citenamefont
  {Navarrete-Benlloch},\ and\ \citenamefont {Hartmann}}]{2016Abdi233604}%
  \BibitemOpen
  \bibfield  {author} {\bibinfo {author} {\bibfnamefont {M.}~\bibnamefont
  {Abdi}}, \bibinfo {author} {\bibfnamefont {P.}~\bibnamefont
  {Degenfeld-Schonburg}}, \bibinfo {author} {\bibfnamefont {M.}~\bibnamefont
  {Sameti}}, \bibinfo {author} {\bibfnamefont {C.}~\bibnamefont
  {Navarrete-Benlloch}}, \ and\ \bibinfo {author} {\bibfnamefont {M.~J.}\
  \bibnamefont {Hartmann}},\ }\bibfield  {title} {\enquote {\bibinfo {title}
  {Dissipative optomechanical preparation of macroscopic quantum superposition
  states},}\ }\href {\doibase 10.1103/PhysRevLett.116.233604} {\bibfield
  {journal} {\bibinfo  {journal} {Phys. Rev. Lett.}\ }\textbf {\bibinfo
  {volume} {116}},\ \bibinfo {pages} {233604} (\bibinfo {year}
  {2016})}\BibitemShut {NoStop}%
\bibitem [{\citenamefont {Kimchi-Schwartz}\ \emph {et~al.}(2016)\citenamefont
  {Kimchi-Schwartz}, \citenamefont {Martin}, \citenamefont {Flurin},
  \citenamefont {Aron}, \citenamefont {Kulkarni}, \citenamefont {Tureci},\ and\
  \citenamefont {Siddiqi}}]{2016Kimchi-Schwartz240503}%
  \BibitemOpen
  \bibfield  {author} {\bibinfo {author} {\bibfnamefont {M.~E.}\ \bibnamefont
  {Kimchi-Schwartz}}, \bibinfo {author} {\bibfnamefont {L.}~\bibnamefont
  {Martin}}, \bibinfo {author} {\bibfnamefont {E.}~\bibnamefont {Flurin}},
  \bibinfo {author} {\bibfnamefont {C.}~\bibnamefont {Aron}}, \bibinfo {author}
  {\bibfnamefont {M.}~\bibnamefont {Kulkarni}}, \bibinfo {author}
  {\bibfnamefont {H.~E.}\ \bibnamefont {Tureci}}, \ and\ \bibinfo {author}
  {\bibfnamefont {I.}~\bibnamefont {Siddiqi}},\ }\bibfield  {title} {\enquote
  {\bibinfo {title} {Stabilizing entanglement via symmetry-selective bath
  engineering in superconducting qubits},}\ }\href {\doibase
  10.1103/PhysRevLett.116.240503} {\bibfield  {journal} {\bibinfo  {journal}
  {Phys. Rev. Lett.}\ }\textbf {\bibinfo {volume} {116}},\ \bibinfo {pages}
  {240503} (\bibinfo {year} {2016})}\BibitemShut {NoStop}%
\bibitem [{\citenamefont {Reiter}\ \emph {et~al.}(2016)\citenamefont {Reiter},
  \citenamefont {Reeb},\ and\ \citenamefont {S{\o}rensen}}]{2016Reiter40501}%
  \BibitemOpen
  \bibfield  {author} {\bibinfo {author} {\bibfnamefont {F.}~\bibnamefont
  {Reiter}}, \bibinfo {author} {\bibfnamefont {D.}~\bibnamefont {Reeb}}, \ and\
  \bibinfo {author} {\bibfnamefont {A.~S.}\ \bibnamefont {S{\o}rensen}},\
  }\bibfield  {title} {\enquote {\bibinfo {title} {Scalable dissipative
  preparation of many-body entanglement},}\ }\href {\doibase
  10.1103/PhysRevLett.117.040501} {\bibfield  {journal} {\bibinfo  {journal}
  {Phys. Rev. Lett.}\ }\textbf {\bibinfo {volume} {117}},\ \bibinfo {pages}
  {040501} (\bibinfo {year} {2016})}\BibitemShut {NoStop}%
\bibitem [{\citenamefont {\v{Z}nidari\v{c}}(2016)}]{2016Znidaric30403}%
  \BibitemOpen
  \bibfield  {author} {\bibinfo {author} {\bibfnamefont {M.}~\bibnamefont
  {\v{Z}nidari\v{c}}},\ }\bibfield  {title} {\enquote {\bibinfo {title}
  {Dissipative remote-state preparation in an interacting medium},}\ }\href
  {\doibase 10.1103/PhysRevLett.116.030403} {\bibfield  {journal} {\bibinfo
  {journal} {Phys. Rev. Lett.}\ }\textbf {\bibinfo {volume} {116}},\ \bibinfo
  {pages} {030403} (\bibinfo {year} {2016})}\BibitemShut {NoStop}%
\bibitem [{\citenamefont {Verstraete}\ \emph {et~al.}(2009)\citenamefont
  {Verstraete}, \citenamefont {Wolf},\ and\ \citenamefont
  {Cirac}}]{2009Verstraete633}%
  \BibitemOpen
  \bibfield  {author} {\bibinfo {author} {\bibfnamefont {F.}~\bibnamefont
  {Verstraete}}, \bibinfo {author} {\bibfnamefont {M.~M.}\ \bibnamefont
  {Wolf}}, \ and\ \bibinfo {author} {\bibfnamefont {J.~I.}\ \bibnamefont
  {Cirac}},\ }\bibfield  {title} {\enquote {\bibinfo {title} {Quantum
  computation and quantum-state engineering driven by dissipation},}\ }\href
  {\doibase 10.1038/NPHYS1342} {\bibfield  {journal} {\bibinfo  {journal} {Nat.
  Phys.}\ }\textbf {\bibinfo {volume} {5}},\ \bibinfo {pages} {633} (\bibinfo
  {year} {2009})}\BibitemShut {NoStop}%
\bibitem [{\citenamefont {Kastoryano}\ \emph {et~al.}(2013)\citenamefont
  {Kastoryano}, \citenamefont {Wolf},\ and\ \citenamefont
  {Eisert}}]{2013Kastoryano110501}%
  \BibitemOpen
  \bibfield  {author} {\bibinfo {author} {\bibfnamefont {M.~J.}\ \bibnamefont
  {Kastoryano}}, \bibinfo {author} {\bibfnamefont {M.~M.}\ \bibnamefont
  {Wolf}}, \ and\ \bibinfo {author} {\bibfnamefont {J.}~\bibnamefont
  {Eisert}},\ }\bibfield  {title} {\enquote {\bibinfo {title} {Precisely timing
  dissipative quantum information processing},}\ }\href {\doibase
  10.1103/PhysRevLett.110.110501} {\bibfield  {journal} {\bibinfo  {journal}
  {Phys. Rev. Lett.}\ }\textbf {\bibinfo {volume} {110}},\ \bibinfo {pages}
  {110501} (\bibinfo {year} {2013})}\BibitemShut {NoStop}%
\bibitem [{\citenamefont {Weimer}\ \emph {et~al.}(2010)\citenamefont {Weimer},
  \citenamefont {M\"{u}ller}, \citenamefont {Lesanovsky}, \citenamefont
  {Zoller},\ and\ \citenamefont {B\"{u}chler}}]{2010Weimer382}%
  \BibitemOpen
  \bibfield  {author} {\bibinfo {author} {\bibfnamefont {H.}~\bibnamefont
  {Weimer}}, \bibinfo {author} {\bibfnamefont {M.}~\bibnamefont {M\"{u}ller}},
  \bibinfo {author} {\bibfnamefont {I.}~\bibnamefont {Lesanovsky}}, \bibinfo
  {author} {\bibfnamefont {P.}~\bibnamefont {Zoller}}, \ and\ \bibinfo {author}
  {\bibfnamefont {H.~P.}\ \bibnamefont {B\"{u}chler}},\ }\bibfield  {title}
  {\enquote {\bibinfo {title} {A rydberg quantum simulator},}\ }\href {\doibase
  10.1038/NPHYS1614} {\bibfield  {journal} {\bibinfo  {journal} {Nat. Phys.}\
  }\textbf {\bibinfo {volume} {6}},\ \bibinfo {pages} {382} (\bibinfo {year}
  {2010})}\BibitemShut {NoStop}%
\bibitem [{\citenamefont {Barreiro}\ \emph {et~al.}(2011)\citenamefont
  {Barreiro}, \citenamefont {M\"{u}ller}, \citenamefont {Schindler},
  \citenamefont {Nigg}, \citenamefont {Monz}, \citenamefont {Chwalla},
  \citenamefont {Hennrich}, \citenamefont {Roos}, \citenamefont {Zoller},\ and\
  \citenamefont {Blatt}}]{2011Barreiro486}%
  \BibitemOpen
  \bibfield  {author} {\bibinfo {author} {\bibfnamefont {J.~T.}\ \bibnamefont
  {Barreiro}}, \bibinfo {author} {\bibfnamefont {M.}~\bibnamefont
  {M\"{u}ller}}, \bibinfo {author} {\bibfnamefont {P.}~\bibnamefont
  {Schindler}}, \bibinfo {author} {\bibfnamefont {D.}~\bibnamefont {Nigg}},
  \bibinfo {author} {\bibfnamefont {T.}~\bibnamefont {Monz}}, \bibinfo {author}
  {\bibfnamefont {M.}~\bibnamefont {Chwalla}}, \bibinfo {author} {\bibfnamefont
  {M.}~\bibnamefont {Hennrich}}, \bibinfo {author} {\bibfnamefont {C.~F.}\
  \bibnamefont {Roos}}, \bibinfo {author} {\bibfnamefont {P.}~\bibnamefont
  {Zoller}}, \ and\ \bibinfo {author} {\bibfnamefont {R.}~\bibnamefont
  {Blatt}},\ }\bibfield  {title} {\enquote {\bibinfo {title} {An open-system
  quantum simulator with trapped ions},}\ }\href {\doibase 10.1038/nature09801}
  {\bibfield  {journal} {\bibinfo  {journal} {Nature (London)}\ }\textbf
  {\bibinfo {volume} {470}},\ \bibinfo {pages} {486} (\bibinfo {year}
  {2011})}\BibitemShut {NoStop}%
\bibitem [{\citenamefont {Aharonov}\ and\ \citenamefont
  {Vaidman}(1993)}]{1993Aharonov38}%
  \BibitemOpen
  \bibfield  {author} {\bibinfo {author} {\bibfnamefont {Y.}~\bibnamefont
  {Aharonov}}\ and\ \bibinfo {author} {\bibfnamefont {L.}~\bibnamefont
  {Vaidman}},\ }\bibfield  {title} {\enquote {\bibinfo {title} {Measurement of
  the schr\"{o}dinger wave of a single particle},}\ }\href {\doibase
  10.1016/0375-9601(93)90724-E} {\bibfield  {journal} {\bibinfo  {journal}
  {Phys. Lett. A}\ }\textbf {\bibinfo {volume} {178}},\ \bibinfo {pages} {38}
  (\bibinfo {year} {1993})}\BibitemShut {NoStop}%
\bibitem [{\citenamefont {Nielsen}\ and\ \citenamefont
  {Chuang}(2010)}]{2010Nielsen}%
  \BibitemOpen
  \bibfield  {author} {\bibinfo {author} {\bibfnamefont {M.~A.}\ \bibnamefont
  {Nielsen}}\ and\ \bibinfo {author} {\bibfnamefont {I.~L.}\ \bibnamefont
  {Chuang}},\ }\href@noop {} {\emph {\bibinfo {title} {Quantum Computation and
  Quantum Information}}}\ (\bibinfo  {publisher} {Cambridge University Press,
  Cambridge, England},\ \bibinfo {year} {2010})\BibitemShut {NoStop}%
\bibitem [{201()}]{2019SM}%
  \BibitemOpen
  \href@noop {} {}\bibinfo {note} {See Supplemental Material at [URL will be
  inserted by publisher] for the discussion on the optimality of the
  conventional strategy, the proof of formula (\ref{2st-main-formula}), and the
  details on the estimation error in the multi-parameter case.}\BibitemShut
  {Stop}%
\bibitem [{\citenamefont {Zanardi}\ and\ \citenamefont
  {Venuti}(2014)}]{2014Zanardi240406}%
  \BibitemOpen
  \bibfield  {author} {\bibinfo {author} {\bibfnamefont {P.}~\bibnamefont
  {Zanardi}}\ and\ \bibinfo {author} {\bibfnamefont {L.~Campos}\ \bibnamefont
  {Venuti}},\ }\bibfield  {title} {\enquote {\bibinfo {title} {Coherent quantum
  dynamics in steady-state manifolds of strongly dissipative systems},}\ }\href
  {\doibase 10.1103/PhysRevLett.113.240406} {\bibfield  {journal} {\bibinfo
  {journal} {Phys. Rev. Lett.}\ }\textbf {\bibinfo {volume} {113}},\ \bibinfo
  {pages} {240406} (\bibinfo {year} {2014})}\BibitemShut {NoStop}%
\bibitem [{1no()}]{1note}%
  \BibitemOpen
  \href@noop {} {}\bibinfo {note} {Unless otherwise stated, the Hilbert-Schmidt
  norm is adopted. That is, for an operator $X$, the norm reads
  $\norm{X}:=\sqrt{\tr(X^\dagger X)}$, while for a superoperator $\mathcal{E}$,
  it is the induced norm defined as $\norm{\mathcal{E}}:=\sup_{\norm{X}\leq
  1}\norm{\mathcal{E}(X)}$.}\BibitemShut {Stop}%
\bibitem [{\citenamefont {Piacentini}\ \emph {et~al.}(2017)\citenamefont
  {Piacentini}, \citenamefont {Avella}, \citenamefont {Rebufello},
  \citenamefont {Lussana}, \citenamefont {Villa}, \citenamefont {Tosi},
  \citenamefont {Gramegna}, \citenamefont {Brida}, \citenamefont {Cohen},
  \citenamefont {Vaidman}, \citenamefont {Degiovanni},\ and\ \citenamefont
  {Genovese}}]{2017Piacentini1191}%
  \BibitemOpen
  \bibfield  {author} {\bibinfo {author} {\bibfnamefont {F.}~\bibnamefont
  {Piacentini}}, \bibinfo {author} {\bibfnamefont {A.}~\bibnamefont {Avella}},
  \bibinfo {author} {\bibfnamefont {E.}~\bibnamefont {Rebufello}}, \bibinfo
  {author} {\bibfnamefont {R.}~\bibnamefont {Lussana}}, \bibinfo {author}
  {\bibfnamefont {F.}~\bibnamefont {Villa}}, \bibinfo {author} {\bibfnamefont
  {A.}~\bibnamefont {Tosi}}, \bibinfo {author} {\bibfnamefont {M.}~\bibnamefont
  {Gramegna}}, \bibinfo {author} {\bibfnamefont {G.}~\bibnamefont {Brida}},
  \bibinfo {author} {\bibfnamefont {E.}~\bibnamefont {Cohen}}, \bibinfo
  {author} {\bibfnamefont {L.}~\bibnamefont {Vaidman}}, \bibinfo {author}
  {\bibfnamefont {I.~P.}\ \bibnamefont {Degiovanni}}, \ and\ \bibinfo {author}
  {\bibfnamefont {M.}~\bibnamefont {Genovese}},\ }\bibfield  {title} {\enquote
  {\bibinfo {title} {Determining the quantum expectation value by measuring a
  single photon},}\ }\href {\doibase 10.1038/NPHYS4223} {\bibfield  {journal}
  {\bibinfo  {journal} {Nat. Phys.}\ }\textbf {\bibinfo {volume} {13}},\
  \bibinfo {pages} {1191} (\bibinfo {year} {2017})}\BibitemShut {NoStop}%
\end{thebibliography}

\begin{thebibliography}{4}%
\makeatletter
\providecommand \@ifxundefined [1]{%
 \@ifx{#1\undefined}
}%
\providecommand \@ifnum [1]{%
 \ifnum #1\expandafter \@firstoftwo
 \else \expandafter \@secondoftwo
 \fi
}%
\providecommand \@ifx [1]{%
 \ifx #1\expandafter \@firstoftwo
 \else \expandafter \@secondoftwo
 \fi
}%
\providecommand \natexlab [1]{#1}%
\providecommand \enquote  [1]{``#1''}%
\providecommand \bibnamefont  [1]{#1}%
\providecommand \bibfnamefont [1]{#1}%
\providecommand \citenamefont [1]{#1}%
\providecommand \href@noop [0]{\@secondoftwo}%
\providecommand \href [0]{\begingroup \@sanitize@url \@href}%
\providecommand \@href[1]{\@@startlink{#1}\@@href}%
\providecommand \@@href[1]{\endgroup#1\@@endlink}%
\providecommand \@sanitize@url [0]{\catcode `\\12\catcode `\$12\catcode
  `\&12\catcode `\#12\catcode `\^12\catcode `\_12\catcode `\%12\relax}%
\providecommand \@@startlink[1]{}%
\providecommand \@@endlink[0]{}%
\providecommand \url  [0]{\begingroup\@sanitize@url \@url }%
\providecommand \@url [1]{\endgroup\@href {#1}{\urlprefix }}%
\providecommand \urlprefix  [0]{URL }%
\providecommand \Eprint [0]{\href }%
\providecommand \doibase [0]{http://dx.doi.org/}%
\providecommand \selectlanguage [0]{\@gobble}%
\providecommand \bibinfo  [0]{\@secondoftwo}%
\providecommand \bibfield  [0]{\@secondoftwo}%
\providecommand \translation [1]{[#1]}%
\providecommand \BibitemOpen [0]{}%
\providecommand \bibitemStop [0]{}%
\providecommand \bibitemNoStop [0]{.\EOS\space}%
\providecommand \EOS [0]{\spacefactor3000\relax}%
\providecommand \BibitemShut  [1]{\csname bibitem#1\endcsname}%
\let\auto@bib@innerbib\@empty
\bibitem [{\citenamefont {Giovannetti}\ \emph {et~al.}(2006)\citenamefont
  {Giovannetti}, \citenamefont {Lloyd},\ and\ \citenamefont
  {Maccone}}]{sm2006Giovannetti10401}%
  \BibitemOpen
  \bibfield  {author} {\bibinfo {author} {\bibfnamefont {V.}~\bibnamefont
  {Giovannetti}}, \bibinfo {author} {\bibfnamefont {S.}~\bibnamefont {Lloyd}},
  \ and\ \bibinfo {author} {\bibfnamefont {L.}~\bibnamefont {Maccone}},\
  }\bibfield  {title} {\enquote {\bibinfo {title} {Quantum metrology},}\ }\href
  {\doibase 10.1103/PhysRevLett.96.010401} {\bibfield  {journal} {\bibinfo
  {journal} {Phys. Rev. Lett.}\ }\textbf {\bibinfo {volume} {96}},\ \bibinfo
  {pages} {010401} (\bibinfo {year} {2006})}\BibitemShut {NoStop}%
\bibitem [{\citenamefont {Braunstein}\ and\ \citenamefont
  {Caves}(1994)}]{sm1994Braunstein3439}%
  \BibitemOpen
  \bibfield  {author} {\bibinfo {author} {\bibfnamefont {S.~L.}\ \bibnamefont
  {Braunstein}}\ and\ \bibinfo {author} {\bibfnamefont {C.~M.}\ \bibnamefont
  {Caves}},\ }\bibfield  {title} {\enquote {\bibinfo {title} {Statistical
  distance and the geometry of quantum states},}\ }\href {\doibase
  10.1103/PhysRevLett.72.3439} {\bibfield  {journal} {\bibinfo  {journal}
  {Phys. Rev. Lett.}\ }\textbf {\bibinfo {volume} {72}},\ \bibinfo {pages}
  {3439} (\bibinfo {year} {1994})}\BibitemShut {NoStop}%
\bibitem [{\citenamefont {Nielsen}\ and\ \citenamefont
  {Chuang}(2010)}]{sm2010Nielsen}%
  \BibitemOpen
  \bibfield  {author} {\bibinfo {author} {\bibfnamefont {M.~A.}\ \bibnamefont
  {Nielsen}}\ and\ \bibinfo {author} {\bibfnamefont {I.~L.}\ \bibnamefont
  {Chuang}},\ }\href@noop {} {\emph {\bibinfo {title} {Quantum Computation and
  Quantum Information}}}\ (\bibinfo  {publisher} {Cambridge University Press,
  Cambridge, England},\ \bibinfo {year} {2010})\BibitemShut {NoStop}%
\bibitem [{\citenamefont {Kato}(1995)}]{sm1995Kato}%
  \BibitemOpen
  \bibfield  {author} {\bibinfo {author} {\bibfnamefont {T.}~\bibnamefont
  {Kato}},\ }\href@noop {} {\emph {\bibinfo {title} {Perturbation Theory for
  Linear Operators}}}\ (\bibinfo  {publisher} {Springer},\ \bibinfo {year}
  {1995})\BibitemShut {NoStop}%
\end{thebibliography}
%

\onecolumngrid
\clearpage

\renewcommand{\theequation}{\thesubsection S.\arabic{equation}}
\setcounter{equation}{0}
\def\Eqf{{8}}

\section*{\large{Supplemental Material}}

\section{optimality of the conventional strategy}

\begin{figure}[htbp]
\includegraphics[width=0.4\textwidth]{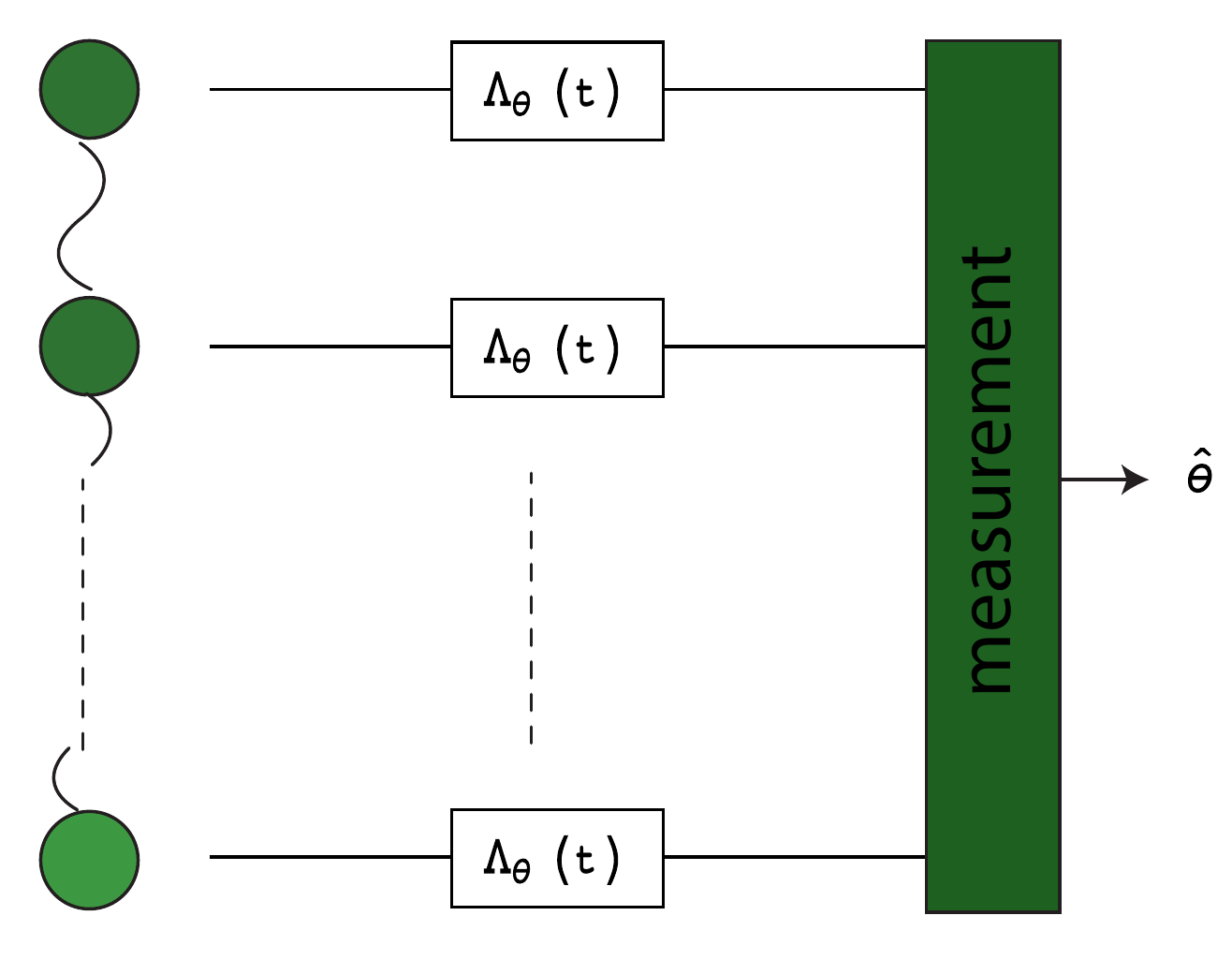}
\caption{General scheme for conventional quantum metrology. $N$ probes, prepared in an initial state, are sent through $N$ parallel channels $\Lambda_\theta(t)$. A measurement is performed on the final state, from which the parameter $\theta$ is estimated via an inference rule $\hat{\theta}$.}
\label{sm-fig1}
\end{figure}

In this section, we show that the strategy presented in the example paragraph of the main text is optimal among conventional strategies of quantum metrology  \cite{sm2006Giovannetti10401}.
Let
\begin{eqnarray}\label{sm-channel}
\Lambda_\theta(t):=e^{\mathcal{L}_\theta t}
\end{eqnarray}
be the quantum channel associated with the generalized amplitude damping process. To estimate the parameter $\theta$ characterizing this channel, a conventional strategy \cite{sm2006Giovannetti10401} is to send $N$ probes through $N$ parallel channels $\Lambda_\theta(t)$, measure them at the output, and use an inference rule $\hat{\theta}(\bm{x})$ to extract the value of $\theta$ from the measurement result $\bm{x}$ (see Fig. \ref{sm-fig1}).
Therefore, a general scheme of conventional quantum metrology consists of three ingredients: an input state of $N$ probes, a measurement at the output, and an inference rule. In the following, we do not impose any restriction on these ingredients. That is, the input state can be an arbitrary, possibly highly entangled, state; the measurement is a general, not necessarily local, POVM; and the inference rule can be biased or unbiased. Additionally, $t$ appearing in Eq.~(\ref{sm-channel}) can take any non-negative
value; it is not necessarily to fulfill the condition assumed in the main text, which requires that the evolution time $t$ is much longer than the relaxation time of the channel.

A well-known measure quantifying the deviation of the estimator $\hat{\theta}(\bm{x})$ from the true value $\theta$ reads
\cite{sm1994Braunstein3439}
\begin{eqnarray}
\Delta\theta(\bm{x}):=\frac{\hat{\theta}(\bm{x})}{\abs{d\expt{\hat{\theta}}/d\theta}}
-\theta.
\end{eqnarray}
Here, $\expt{\hat{\theta}}$ denotes the statistical average of $\hat{\theta}(\bm{x})$ over potential outcomes $\bm{x}$. Accordingly, the estimation error can be defined as \cite{sm1994Braunstein3439}
\begin{eqnarray}
\delta\theta:=\sqrt{\expt{\Delta\theta^2}},
\end{eqnarray}
i.e., the square root of the statistical average of $\Delta\theta^2(\bm{x})$ over potential outcomes $\bm{x}$. In particular, if $\hat{\theta}(\bm{x})$ is unbiased, i.e., $\expt{\hat{\theta}}=\theta$, there is
\begin{eqnarray}
\Delta\theta(\bm{x})=\hat{\theta}(\bm{x})-\expt{\hat{\theta}},
\end{eqnarray}
indicating that $\delta\theta$ is simply the standard deviation,
\begin{eqnarray}\label{sec1:df-error}
\delta\theta=\sqrt{\textrm{Var}(\hat{\theta})},
\end{eqnarray}
for an unbiased estimator $\hat{\theta}(\bm{x})$. This fact has been used in the main text.

Using the quantum Cram\'{e}r-Rao inequality \cite{sm1994Braunstein3439}, we have that the estimation error $\delta\theta$ is lower bounded by
\begin{eqnarray}\label{Q-CR}
\delta\theta\geq\frac{1}{\sqrt{F\left[\Lambda_\theta^{\otimes N}(t)[\rho(0)]\right]}}.
\end{eqnarray}
Here, $F$ is the quantum Fisher information (QFI) and $\Lambda_\theta^{\otimes N}(t)[\rho(0)]$ is the output state of the $N$ probes, where $\rho(0)$ denotes the input state. To evaluate the quantity $F\left[\Lambda_\theta^{\otimes N}(t)[\rho(0)]\right]$, we
introduce two amplitude damping channels
\cite{sm2010Nielsen}, $\Lambda_i(t)$, $i=0,1$, transforming the Bloch vector as
\begin{eqnarray}
\Lambda_0(t):\ \ (r_x,r_y,r_z)\rightarrow(e^{-\frac{t}{2}}r_x,
e^{-\frac{t}{2}}r_y,1-e^{-t}+e^{-t}r_z),
\end{eqnarray}
and
\begin{eqnarray}
\Lambda_1(t):\ \ (r_x,r_y,r_z)\rightarrow(e^{-\frac{t}{2}}r_x,
e^{-\frac{t}{2}}r_y,e^{-t}-1+e^{-t}r_z),
\end{eqnarray}
respectively.
Noting that the effect of $\Lambda_\theta(t)$ is
\begin{eqnarray}
(r_x,r_y,r_z)\rightarrow
\left(e^{-\frac{t}{2}}r_x,
e^{-\frac{t}{2}}r_y,(2\theta-1)(1-e^{-t})+e^{-t}r_z\right),
\end{eqnarray}
we have
\begin{eqnarray}\label{eq:decom}
\Lambda_\theta(t)=\theta\Lambda_0(t)+(1-\theta)\Lambda_1(t).
\end{eqnarray}
Equation (\ref{eq:decom}) enables us to rewrite $\Lambda_\theta(t)$ as a $\theta$-independent quantum channel acting on a larger input space,
\begin{eqnarray}\label{eq:big-map}
\Lambda_\theta(t)[\rho]=\Phi(t)[\rho\otimes\rho_\theta].
\end{eqnarray}
Here, $\rho_\theta=\textrm{diag}(\theta,1-\theta)$ is the steady state, and $\Phi(t)$ is defined as
\begin{eqnarray}
\Phi(t)[\rho\otimes\sigma]:=\sum_{i=0,1}\Lambda_i(t)\otimes\mathcal{E}_i
[\rho\otimes\sigma]
=\sum_{i=0,1}\Lambda_i(t)[\rho]\otimes\mathcal{E}_i
[\sigma],
\end{eqnarray}
where $\mathcal{E}_i[\sigma]:=\bra{i}\sigma\ket{i}$, $i=0,1$.
Using Eq.~(\ref{eq:big-map}), we have
\begin{eqnarray}
F\left[\Lambda_\theta^{\otimes N}(t)[\rho(0)]\right]=
F\left[\Phi^{\otimes N}(t)[\rho(0)\otimes\rho_\theta^{\otimes N}]\right]
\leq F\left[\rho(0)\otimes\rho_\theta^{\otimes N}\right]
=F\left[\rho_\theta^{\otimes N}\right],
\end{eqnarray}
where we have used the monotonicity of the QFI under parameter-independent quantum channels \cite{sm1994Braunstein3439}. Noting that $F\left[\rho_\theta^{\otimes N}\right]=NF\left[\rho_\theta\right]$ and $F[\rho_\theta]=\frac{1}{\theta(1-\theta)}$, we further have
\begin{eqnarray}\label{eq:bound-F}
F\left[\Lambda_\theta^{\otimes N}(t)[\rho(0)]\right]\leq\frac{N}{\theta(1-\theta)}.
\end{eqnarray}
Substituting Eq.~(\ref{eq:bound-F}) into Eq.~(\ref{Q-CR}) yields
\begin{eqnarray}
\delta\theta\geq\sqrt{\frac{\theta(1-\theta)}{N}},
\end{eqnarray}
indicating that the strategy presented in the main text is optimal among conventional strategies of quantum metrology.

\section{Proof of formula (\Eqf)}

In this section, we present a proof for formula (\Eqf) in the main text. Note that $\hat{\theta}(q)=f^{-1}(q/N)$ is an unbiased estimator.
By definition,
\begin{eqnarray}
\delta\theta=\sqrt{\textrm{Var}(\hat{\theta})}.
\end{eqnarray}
Using error propagation theory, we have
\begin{eqnarray}\label{delta}
\delta\theta=\sqrt{\textrm{Var}(q)}/\left(N\frac{\partial f}{\partial\theta}\right).
\end{eqnarray}
Here,
\begin{eqnarray}\label{sec2:var-q}
\textrm{Var}(q)
=\int dq\left(q-N\expt{A}_\theta\right)^2\textrm{Pr}(q),
\end{eqnarray}
where
\begin{eqnarray}\label{sec2:prob}
\textrm{Pr}(q):=\bra{q}\tr_\mathscr{S}\mathcal{E}_T^N
(\rho_\theta\otimes\ket{\phi}\bra{\phi})\ket{q}
\end{eqnarray}
denotes the probability distribution of the pointer reading $q$.
Noting that $\ket{\phi}=\int dp \phi(p)\ket{p}$ and $\mathcal{E}_T^N
(\rho_\theta\otimes\ket{p}\bra{p^\prime})=\left(e^{\mathcal{L}_{p,p^\prime}NT}\rho_\theta
\right)\otimes\ket{p}\bra{p^\prime}$, we can rewrite Eq.~(\ref{sec2:prob}) as
\begin{eqnarray}\label{sec2:prob-new}
\textrm{Pr}(q)=
\frac{1}{2\pi}\iint
dpdp^\prime\phi(p)\phi^*(p^\prime)
e^{i(p-p^\prime)q}\tr_\mathscr{S}(e^{\mathcal{L}_{p,p^\prime}NT}\rho_\theta).
\end{eqnarray}

To compute $\textrm{Pr}(q)$, we need to find an expression for the term $\tr_\mathscr{S}(e^{\mathcal{L}_{p,p^\prime}NT}\rho_\theta)$ appearing in Eq.~(\ref{sec2:prob-new}). Noting that
\begin{eqnarray}
\mathcal{L}_{p,p^\prime}=\mathcal{L}_\theta+T^{-1}\mathcal{K},
\end{eqnarray}
with
\begin{eqnarray}
\mathcal{K}\rho:=-i\left(pA\rho-p^\prime\rho A\right),
\end{eqnarray}
we can interpret $\mathcal{L}_{p,p^\prime}$ as the sum of the ``unperturbed term''  $\mathcal{L}_\theta$ and the perturbation term $\mathcal{K}$. Since $\rho_\theta$ is an eigenstate of the unperturbed term $\mathcal{L}_\theta$, i.e., $\mathcal{L}_\theta\rho_\theta=0$, it must be an approximate eigenstate of $\mathcal{L}_{p,p^\prime}$. Perturbation theory \cite{sm1995Kato} tells us that the difference between such an approximate eigenstate and the associated true eigenstate of $\mathcal{L}_{p,p^\prime}$ is of the order $O(1/T)$. Using this fact as well as noting that $\norm{e^{\mathcal{L}_{p,p^\prime}NT}}=O(1)$, we have
\begin{eqnarray}\label{sec2:Lpp}
\tr_\mathscr{S}(e^{\mathcal{L}_{p,p^\prime}NT}\rho_\theta)=\tr_\mathscr{S} (e^{\lambda NT}\rho_\theta)=e^{\lambda NT},
\end{eqnarray}
up to a term of order $O(1/T)$, which is negligible as $T\gg 1$.
Here, $\lambda$ denotes the corresponding eigenvalue of $\mathcal{L}_{p,p^\prime}$, which can be expressed as a series
\begin{eqnarray}\label{sec2:eigv}
\lambda=\lambda^{(0)}+T^{-1}\lambda^{(1)}+T^{-2}\lambda^{(2)}+\cdots,
\end{eqnarray}
where $\lambda^{(n)}$ denotes its $n$-th order perturbation.
Substituting Eq.~(\ref{sec2:eigv}) into Eq.~(\ref{sec2:Lpp}) gives
\begin{eqnarray}\label{sec2:Lpp2}
\tr_\mathscr{S}(e^{\mathcal{L}_{p,p^\prime}NT}\rho_\theta)=e^{\lambda^{(0)}NT+\lambda^{(1)}N+
\lambda^{(2)}N/T}.
\end{eqnarray}
Here, we have ignored terms of the order $O(N/T^2)$, which are negligible because of $N\leq N_\textrm{max}:=O(T)$, a condition that has been assumed in the main text.

According to perturbation theory \cite{sm1995Kato}, there are
\begin{eqnarray}\label{sec2:0th}
\lambda^{(0)}=0,
\end{eqnarray}
\begin{eqnarray}\label{sec2:1st}
\lambda^{(1)}=\tr_\mathscr{S}(\mathcal{K}\rho_\theta)
=-i(p-p^\prime)\expt{A}_\theta,
\end{eqnarray}
and
\begin{eqnarray}\label{sec2:2nd}
\lambda^{(2)}=-\tr_\mathscr{S}\left[\mathcal{K}\mathcal{S_\theta}\mathcal{K}(\rho_\theta)\right]
=p(p-p^\prime)\tr_\mathscr{S}[A\mathcal{S}_\theta(A\rho_\theta)]-p^\prime(p-p^\prime)
\tr_\mathscr{S}[A\mathcal{S}_\theta(\rho_\theta A)],
\label{secondorder}
\end{eqnarray}
with $\mathcal{S}_\theta$ being the pseudoinverse of $\mathcal{L}_\theta$, as defined in the main text. Here, we have used the formulae dealing with perturbations of eigenvalues of linear operators (see page 79 of Ref.~\cite{sm1995Kato}). Note that the original formulae in Ref.~\cite{sm1995Kato} are expressed in terms of linear operators, but here we have reformulated them in terms of superoperators for serving our purpose.

Noting that $\mathcal{S}_\theta$ is a Hermitian map, i.e.,
\begin{eqnarray}
\mathcal{S}_\theta(X)^\dagger=\mathcal{S}_\theta(X^\dagger),
\end{eqnarray}
we have
\begin{eqnarray}\label{sec2:complex-relation}
\tr_\mathscr{S}[A\mathcal{S}_\theta(A\rho_\theta)]=\tr_\mathscr{S}[A\mathcal{S}_\theta(\rho_\theta A)]^*.
\end{eqnarray}
Using Eq.~(\ref{sec2:complex-relation}), we can rewrite Eq.~(\ref{sec2:2nd}) as
\begin{eqnarray}\label{sec2:2nd-new}
\lambda^{(2)}=(p-p^\prime)^2\textrm{Re}\left(\tr_\mathscr{S}
[A\mathcal{S}_\theta(A\rho_\theta)]\right)
+
i(p-p^\prime)(p+p^\prime)\textrm{Im}\left(\tr_\mathscr{S}[A\mathcal{S}_\theta(A\rho_\theta)]\right).
\end{eqnarray}
Inserting Eqs.~(\ref{sec2:0th}), (\ref{sec2:1st}), and (\ref{sec2:2nd-new}) into Eq.~(\ref{sec2:Lpp2}), we arrive at the desired expression
\begin{eqnarray}\label{sec2:Lpp-final}
\tr_\mathscr{S}(e^{\mathcal{L}_{p,p^\prime}NT}\rho_\theta)=
e^{-i(p-p^\prime)N\expt{A}_\theta +\frac{N}{T}(p-p^\prime)^2\textrm{Re}\left(\tr_\mathscr{S}
[A\mathcal{S}_\theta(A\rho_\theta)]\right)+
\frac{iN}{T}(p-p^\prime)(p+p^\prime)\textrm{Im}\left(
\tr_\mathscr{S}[A\mathcal{S}_\theta(A\rho_\theta)]\right)}.
\end{eqnarray}

As assumed in the main text, the coordinate representation of $\ket{\phi}$ is a Gaussian with standard deviation $\sigma$. Hence, there is
\begin{eqnarray}
\phi(p)=\frac{1}{(2\pi\sigma^{\prime2})^{1/4}}e^{-\frac{p^2}{4\sigma^{\prime 2}}},
\end{eqnarray}
where
\begin{eqnarray}
\sigma^\prime=\frac{1}{2\sigma}.
\end{eqnarray}
Using the above expression of $\phi(p)$ and substituting Eq.~(\ref{sec2:Lpp-final}) into Eq.~(\ref{sec2:prob-new}),
we have
\begin{eqnarray}\label{sec2:Pr}
\textrm{Pr}(q)=\frac{1}{2\pi\sqrt{2\pi\sigma^{\prime 2}}}
\iint dpdp^\prime e^{-\frac{p^2+p^{\prime 2}}{4\sigma^{\prime 2}}} e^{i(p-p^\prime)(q-N\expt{A}_\theta)+
\frac{N}{T}(p-p^\prime)^2\textrm{Re}\left(\tr_\mathscr{S}
[A\mathcal{S}_\theta(A\rho_\theta)]\right)+
\frac{iN}{T}(p-p^\prime)(p+p^\prime)\textrm{Im}
\left(\tr_\mathscr{S}[A\mathcal{S}_\theta(A\rho_\theta)]\right)}.\nonumber\\
\end{eqnarray}
Inserting Eq.~(\ref{sec2:Pr}) into Eq.~(\ref{sec2:var-q}) and simplifying the resultant equation by defining new variables
\begin{eqnarray}
x:&=&p-p^\prime,\nonumber\\
y:&=&p+p^\prime,
\end{eqnarray}
we obtain
\begin{eqnarray}\label{sec2:vq-new}
\textrm{Var}(q)=\frac{1}{4\pi\sqrt{2\pi\sigma^{\prime 2}}}\int dq
(q-N\expt{A}_\theta)^2\int dx
e^{-\frac{x^2}{8\sigma^{\prime 2}}}
e^{ix(q-N\expt{A}_\theta)}
e^{\frac{N}{T}x^2\textrm{Re}\left(\tr_\mathscr{S}
[A\mathcal{S}_\theta(A\rho_\theta)]\right)}
\int dy e^{-\frac{y^2}{8\sigma^{\prime 2}}}
e^{\frac{iN}{T}xy\textrm{Im}\left(\tr_\mathscr{S}
[A\mathcal{S}_\theta(A\rho_\theta)]\right)}.\nonumber\\
\end{eqnarray}
Here, the fact $dpdp^\prime=\frac{1}{2}dxdy$ has been used.

Expanding terms $e^{\frac{N}{T}x^2\textrm{Re}\left(\tr_\mathscr{S}[A\mathcal{S}_\theta(A\rho_\theta)]\right)}$ and $e^{\frac{iN}{T}xy\textrm{Im}\left(\tr_\mathscr{S}[A\mathcal{S}_\theta(A\rho_\theta)]\right)}$ appearing in Eq.~(\ref{sec2:vq-new}) as power series
\begin{eqnarray}
e^{\frac{N}{T}x^2\textrm{Re}\left(\tr_\mathscr{S}
[A\mathcal{S}_\theta(A\rho_\theta)]\right)}
&=&\sum_{m=0}^\infty\frac{\left[\frac{N}{T}\textrm{Re}
\left(\tr_\mathscr{S}[A\mathcal{S}_\theta(A\rho_\theta)]\right)
\right]^m}{m!}x^{2m},\nonumber\\
e^{\frac{iN}{T}xy\textrm{Im}\left(\tr_\mathscr{S}
[A\mathcal{S}_\theta(A\rho_\theta)]\right)}
&=&\sum_{n=0}^\infty \frac{\left[\frac{iN}{T}\textrm{Im}
\left(\tr_\textrm{S}[A\mathcal{S}_\theta(A\rho_\theta)]\right)\right]^n}{n!}
x^ny^n,
\end{eqnarray}
we have
\begin{eqnarray}\label{vq-new}
\textrm{Var}(q)=\sum_{m=0}^\infty\sum_{n=0}^\infty
\frac{\left[\frac{N}{T}\textrm{Re}\left(\tr_\mathscr{S}
[A\mathcal{S}_\theta(A\rho_\theta)]\right)
\right]^m}{m!}
\frac{\left[\frac{iN}{T}\textrm{Im}\left(\tr_\mathscr{S}[A\mathcal{S}_\theta(A\rho_\theta)]\right)
\right]^n}{n!}
F(m,n),
\end{eqnarray}
where
\begin{eqnarray}
F(m,n):=\frac{1}{4\pi\sqrt{2\pi\sigma^{\prime 2}}}\int dq
(q-N\expt{A}_\theta)^2\int dx
e^{-\frac{x^2}{8\sigma^{\prime 2}}}
e^{ix(q-N\expt{A}_\theta)}x^{2m+n}\int dy
e^{-\frac{y^2}{8\sigma^{\prime 2}}}y^{n}.
\end{eqnarray}
Since
\begin{eqnarray}
\int dy e^{-\frac{y^2}{8\sigma^{\prime 2}}} y^n=0,\quad n\in\textrm{odd},
\end{eqnarray}
there is
\begin{eqnarray}
F(m,n)=0, \quad n\in\textrm{odd}.
\end{eqnarray}
Further, since
\begin{eqnarray}
\int dq(q-N\expt{A}_\theta)^2\int dx e^{-\frac{x^2}{8\sigma^{\prime 2}}} e^{ix(q-N\expt{A}_\theta)} x^{2l}=0, \quad l\geq 2,
\end{eqnarray}
there is
\begin{eqnarray}
F(m,n)=0, \quad 2m+n=4,6,8,10,\cdots.
\end{eqnarray}
So, the non-vanishing terms are $F(0,0)$, $F(1,0)$, and $F(0,2)$, given by
\begin{eqnarray}\label{sec2:F}
F(0,0)=\sigma^2,\quad
F(1,0)=-2, \quad\textrm{and}\quad
F(0,2)=-2/\sigma^2,
\end{eqnarray}
respectively. Inserting Eq.~(\ref{sec2:F}) into Eq.~(\ref{vq-new}) yields
\begin{eqnarray}\label{sec2:vq-f}
\textrm{Var}(q)=\sigma^2-\frac{2N\textrm{Re}\left(\tr_\mathscr{S}
[A\mathcal{S}_\theta(A\rho_\theta)]
\right)}{T}
+\left[\frac{N\textrm{Im}\left(\tr_\mathscr{S}
[A\mathcal{S}_\theta(A\rho_\theta)]\right)}{T\sigma }\right]^2.
\end{eqnarray}
Finally, substituting Eq.~(\ref{sec2:vq-f}) into Eq.~(\ref{delta}), we obtain
\begin{eqnarray}\label{sec2:formula}
\delta\theta=\sqrt{\sigma^2-\frac{2N\textrm{Re}\left(
\tr_\mathscr{S}[A\mathcal{S}_\theta(A\rho_\theta)]\right)}{T}
+\left[\frac{N\textrm{Im}\left(\tr_\mathscr{S}[A\mathcal{S}_\theta(A\rho_\theta)]\right)}{T\sigma }\right]^2}/\left(N\frac{\partial f}{\partial\theta}\right).
\end{eqnarray}
This completes the proof.

\section{The estimation error in the multi-parameter case}

In this section, we address the estimation error of our approach in the multi-parameter case. To quantify the estimation error in this case, we adopt the following measure,
\begin{eqnarray}\label{sec3:df-error}
\delta\bm{\theta}:=\sqrt{\int d\bm{q} \norm{\hat{\bm{\theta}}(\bm{q})-\bm{\theta}}^2\textrm{Pr}(\bm{q})}.
\end{eqnarray}
Here, $d\bm{q}:=dq_1\cdots dq_M$, $\norm{\hat{\bm{\theta}}(\bm{q})-\bm{\theta}}^2:=
\sum_i\left(\hat{\theta}_i(\bm{q})-\theta_i\right)^2$, and $\textrm{Pr}(\bm{q}):=\textrm{Pr}(q_1)\cdots\textrm{Pr}(q_M)$, where $\textrm{Pr}(q_i)$ denotes the probability distribution of the pointer reading
$q_i$. The above measure is a natural generalization of the measure defined in Ref.~\cite{sm1994Braunstein3439}, as Eq.~(\ref{sec3:df-error}) reduces to Eq.~(\ref{sec1:df-error}) in the single-parameter case.


Using Taylor-series expansion of $\bm{f}^{-1}(\bm{q}/N)$ and noting that $\bm{\theta}=\bm{f}^{-1}(\expt{A_1}_\theta,\cdots,\expt{A_M}_\theta)$, we have
\begin{eqnarray}\label{sec3:ts}
\norm{\hat{\bm{\theta}}(\bm{q})-\bm{\theta}}^2=
\norm{\bm{J}_{\bm{f}^{-1}}\left(\bm{q}/N-(\expt{A_1}_\theta,\cdots,\expt{A_M}_\theta)
\right)^T}^2
=\frac{1}{N^2}\sum_{i}\left[\sum_j(\bm{J}_{\bm{f}^{-1}})_{ij}(q_j-N\expt{A_j}_\theta)\right]^2.
\end{eqnarray}
Here, $\bm{J}_{\bm{f}^{-1}}$ is the Jacobian matrix associated with $\bm{f}^{-1}$, with $(\bm{J}_{\bm{f}^{-1}})_{ij}$ denoting its $ij$-th element. Equation (\ref{sec3:ts}) holds for the $\bm{q}/N$ that is close to $(\expt{A_1}_\theta,\cdots,\expt{A_M}_\theta)$. Noting that $\textrm{Pr}(\bm{q})$ exponentially decreases to zero in the course of moving $\bm{q}/N$ away from $(\expt{A_1}_\theta,\cdots,\expt{A_M}_\theta)$, we can substitute Eq.~(\ref{sec3:ts}) into Eq.~(\ref{sec3:df-error}), and obtain
\begin{eqnarray}\label{sec3:st1}
\delta\bm{\theta}&=&\frac{1}{N}\sqrt{\int d\bm{q} \sum_i\left[\sum_j(\bm{J}_{\bm{f}^{-1}})_{ij}(q_j-N\expt{A_j}_\theta)\right]^2
\textrm{Pr}(\bm{q})}\nonumber\\
&=&
\frac{1}{N}\sqrt{\sum_{ijk}(\bm{J}_{\bm{f}^{-1}})_{ij}(\bm{J}_{\bm{f}^{-1}})_{ik}\iint dq_jdq_k(q_j-N\expt{A_j}_\theta)
(q_k-N\expt{A_k}_\theta)
\textrm{Pr}(q_j)\textrm{Pr}(q_k)}.
\end{eqnarray}
Simplifying Eq.~(\ref{sec3:st1}) by noting that
\begin{eqnarray}
\int dq_j (q_j-N\expt{A_j}_\theta)\textrm{Pr}(q_j)=0,
\end{eqnarray}
we have
\begin{eqnarray}\label{sec3:st3}
\delta\bm{\theta}=\frac{1}{N}\sqrt{\sum_{ij}(\bm{J}_{\bm{f}^{-1}})_{ij}^2
\textrm{Var}(q_j)}.
\end{eqnarray}
Substituting Eq.~(\ref{sec2:vq-f}) into Eq.~(\ref{sec3:st3}), we reach the formula describing the error $\delta\bm{\theta}$,
\begin{eqnarray}\label{sec3:formula}
\delta\bm{\theta}=\frac{1}{N}\sqrt{\sum_{ij}(\bm{J}_{\bm{f}^{-1}})_{ij}^2
\left[\sigma^2-\frac{2N\textrm{Re}\left(\tr_\mathscr{S}
[A_j\mathcal{S}_\theta(A_j\rho_\theta)]
\right)}{T}
+\frac{N^2\textrm{Im}\left(\tr_\mathscr{S}
[A_j\mathcal{S}_\theta(A_j\rho_\theta)]\right)^2}{T^2\sigma^2}\right]}.
\end{eqnarray}

As can be seen from this formula, as long as $N\leq N_\textrm{max}:=O(T)$, $\delta\bm{\theta}\thicksim 1/N$, that is, our approach gives the Heisenberg
scaling of precision in the multi-parameter case as well. On the other hand, noting that $\bm{J}_{\bm{f}^{-1}}=1/\frac{\partial f}{\partial\theta}$ in the single-parameter case, we deduce that formula (\ref{sec3:formula}) reduces to formula (\ref{sec2:formula}).

\end{document}